\documentclass[
prb
 ,amsmath,amssymb
 ,twocolumn,a4paper,twoside,
 floatfix
]{revtex4-2}
\usepackage[english]{babel}
\usepackage{graphicx}
\usepackage{placeins} 
\usepackage{enumitem} 
\usepackage{dcolumn}
\usepackage{bm}
\usepackage[mathlines,pagewise,modulo]{lineno}
\usepackage[normalem]{ulem}
\usepackage{xcolor}
\definecolor{eloicyan}{rgb}{0.,0.64,0.84}
\definecolor{darkgreen}{rgb}{0.,0.46,0.46}
\definecolor{darkblue}{rgb}{0.1,0.2,0.46}
\definecolor{lightred}{rgb}{1.0,0.3,0.3}
\definecolor{darkred}{rgb}{.8,0.1,0.0}

\newcommand\redout{\bgroup\markoverwith{\textcolor{lightred}{\rule[0.5ex]{2pt}{.7pt}}}\ULon}

\newcommand\cyanout{\bgroup\markoverwith{\textcolor{eloicyan}{\rule[0.5ex]{2pt}{.7pt}}}\ULon}

\newcommand\Lilacout{\bgroup\markoverwith{\textcolor{lilac}{\rule[0.5ex]{2pt}{.7pt}}}\ULon}

\definecolor{lilac}{rgb}{0.6,0.2,0.7}

\newcommand*{\eff}{\rm{eff}}
\newcommand*\dmU{\delta_{\vec{m}}U}

\let\oldvec\vec 
\renewcommand{\vec}[1]{\oldvec{\bm #1}} 
\let\oldhat\hat 
\renewcommand{\hat}[1]{\oldhat{\bm #1}} 
\DeclareMathAlphabet\mathbfcal{OMS}{cmsy}{b}{n} 
\DeclareUnicodeCharacter{2212}{-} 
\renewcommand{\arraystretch}{1.2} 

\begin{document}


\title{Domain wall dynamics in antiferromagnetically-coupled double-lattice systems}

\author{Eloi Haltz}
\affiliation{
Université Paris-Saclay, CNRS, Laboratoire de Physique des Solides, 91405 Orsay, France
}
\affiliation{
School of Physics and Astronomy, University of Leeds, Leeds LS2 9JT, United Kingdom
}
\author{Sachin Krishnia}
\author{Léo Berges}
\author{Alexandra Mougin}
\author{João Sampaio}
 \email{joao.sampaio@universite-paris-saclay.fr}
\affiliation{
Université Paris-Saclay, CNRS, Laboratoire de Physique des Solides, 91405 Orsay, France
}

\date{\today}

\begin{abstract}
    In ferromagnetic materials, the rich dynamics of magnetic domain walls (DWs) under magnetic field or current have been successfully described using the well-known \textit{q}-$\varphi$ analytical model. We demonstrate here that this simple unidimensional model holds for multiple-sublattice materials such as ferrimagnetic alloys or synthetic antiferromagnets (SAF) by using effective parameters, and is in excellent agreement with double-lattice micromagnetic simulations. We obtain analytical laws for the DW velocity and internal precession angle as a function of net magnetisation for different driving forces (magnetic field, spin transfer and spin-orbit torques) and different propagation regimes in ferrimagnetic alloys and SAFs. The model predicts that several distinctive dynamical features occur near or at the magnetic and the angular compensation points when the net magnetization or the net angular momentum of the system vanishes, and we discuss the experimental observations that have been reported for some of them.
    Using a higher degree-of-freedom analytical model that accounts for inter-sublattice distortions, we give analytical expressions for these distortions that agree with the micromagnetic simulations. This model shows that the DW velocity and precession rate are independent of the strength of the inter-sublattice exchange coupling, and justifies the use of the simpler effective parameters model.
\end{abstract}

\maketitle

\section*{Contents}
\begin{enumerate}[label=\Roman*., itemsep=0pt, topsep=0pt,nosep]
    \item Introduction
    \item Modelling DW Dynamics in Multi-Lattice Systems
    \item DW Dynamics
    \begin{enumerate}[label=\Alph*., itemsep=0pt, topsep=0pt]
        \item Material Parameters
        \item DW Driven by Field
        \item DW Driven by STT
        \item DW Driven by SHE
    \end{enumerate}
    \item Effects of Finite Coupling
    \item Conclusion
\end{enumerate}

\section{Introduction}

Antiferromagnetically-coupled two lattice magnetic systems, such as antiferromagnets, ferrimagnets, or synthetic antiferromagnets (SAFs), offer a promising path towards much faster, denser, more robust and efficient spintronic devices~\cite{Manchon2019}.

Their decreased net magnetisation reduces the vulnerability to external fields and to cross-talk from adjacent devices, allowing for higher storage density and increased robustness. Secondly, the reduced net angular momentum leads to drastically faster magnetic dynamics. Promising theoretical results and experimental observations have been recently obtained in such multi-lattice magnetic systems~\cite{Avci2019,Jungwirth2016}, multi-layers architectures~\cite{Yang2015,Hrabec2018,Hrabec2017}, or synthetic alloys~\cite{Kim2017b,Siddiqui2018,Caretta2018,Hirata2019,Woo2018,Haltz2019,Haltz2020}. In particular, it has been shown that the antiferromagnetic coupling between sub-lattices brings the characteristic frequency of magnetisation dynamics up to the THz range~\cite{Shiino2016,Keffer1952, Stremoukhov2019, Oh2017,Lepadatu2017}.

Antiferromagnets are hard to probe and manipulate experimentally as the two sub-lattices are of the same nature and so perfectly balanced. An interesting alternative are systems which mix two distinct sub-lattices, allowing for a selectivity in both the reading and the manipulation of the antiferromagnetic order.
Two very promising and versatile classes of such systems are studied: one where the sublattices are spatially merged and consist of two different chemical species (Fig.~\ref{fig:2DWferri}a), such as rare earth/transition metal ferrimagnetic alloys (RE-TM), and another where the two sublattices are spatially-separated (Fig.~\ref{fig:2DWferri}b), such as in synthetic antiferromagnets (SAF) where two magnetic films are antiferromagnetically coupled through RKKY interaction~\cite{Duine2018,Parkin1990,Parkin1991}.
In both systems, it is possible to change the balance between the moments of the sublattices by changing the composition or temperature of the ferrimagnetic alloy~\cite{Hansen1989}  or the thickness of the layers of the SAF~\cite{Parkin1990}. Two particular configurations are of special interest: the magnetic compensation point (MCP) where the the net magnetisation of the system vanishes, and the  angular compensation point (ACP), where the net angular momentum vanishes. Depending on the material, these compensation points are distinct if the two sub-lattices are chemically different~\cite{Hirata2018} or coincide if they have the same nature.

These systems are already well-known for their interesting static properties \cite{Parkin1991,Hansen1989}.
More recently, the study of dynamics in such systems has produced many interesting results~\cite{Manchon2019}. However, the dynamics of magnetic textures in these systems still lacks a clear, well-understood and straightforward description. Experimental results are often subject to apparently conflicting interpretations, particularly near the compensation points, where the dynamics differ the most from the well-studied ferromagnetic case. \\

In this article, the dynamics of double-sub-lattice films are analytically and numerically investigated through the prism of magnetic domain wall (DW) dynamics. In section \ref{sect:Models}, we discuss different models for DW dynamics.
In section \ref{sect:DW_Dynamics}, we give analytical laws and we evaluate their validity for the DW dynamics under different driving forces: magnetic field, spin transfer torque (STT), and interfacial spin orbit torques (SOT). Finally, we consider the effects of finite coupling in section \ref{sect:Low_Coupling}. For the reader's convenience, all used symbols and their definitions are listed in Table~\ref{tab:glossary}.
We have chosen to use the material parameters of two example real systems: the RE-TM ferrimagnetic alloy CoGd, and the SAF CoFeB/Ru/CoFeB (Fig.~\ref{fig:2DWferri}a and b). These two systems illustrate the cases of spatially merged (CoGd) and segregated (SAF) antiferromagnetically coupled sub-lattices, as well as of systems with distinct (CoGd) and coinciding (SAF) compensation points.

\section{Modelling DW dynamics in multi-lattice systems} \label{sect:Models}

\begin{figure}[ht]
	\centering
	\includegraphics[width=1\columnwidth]{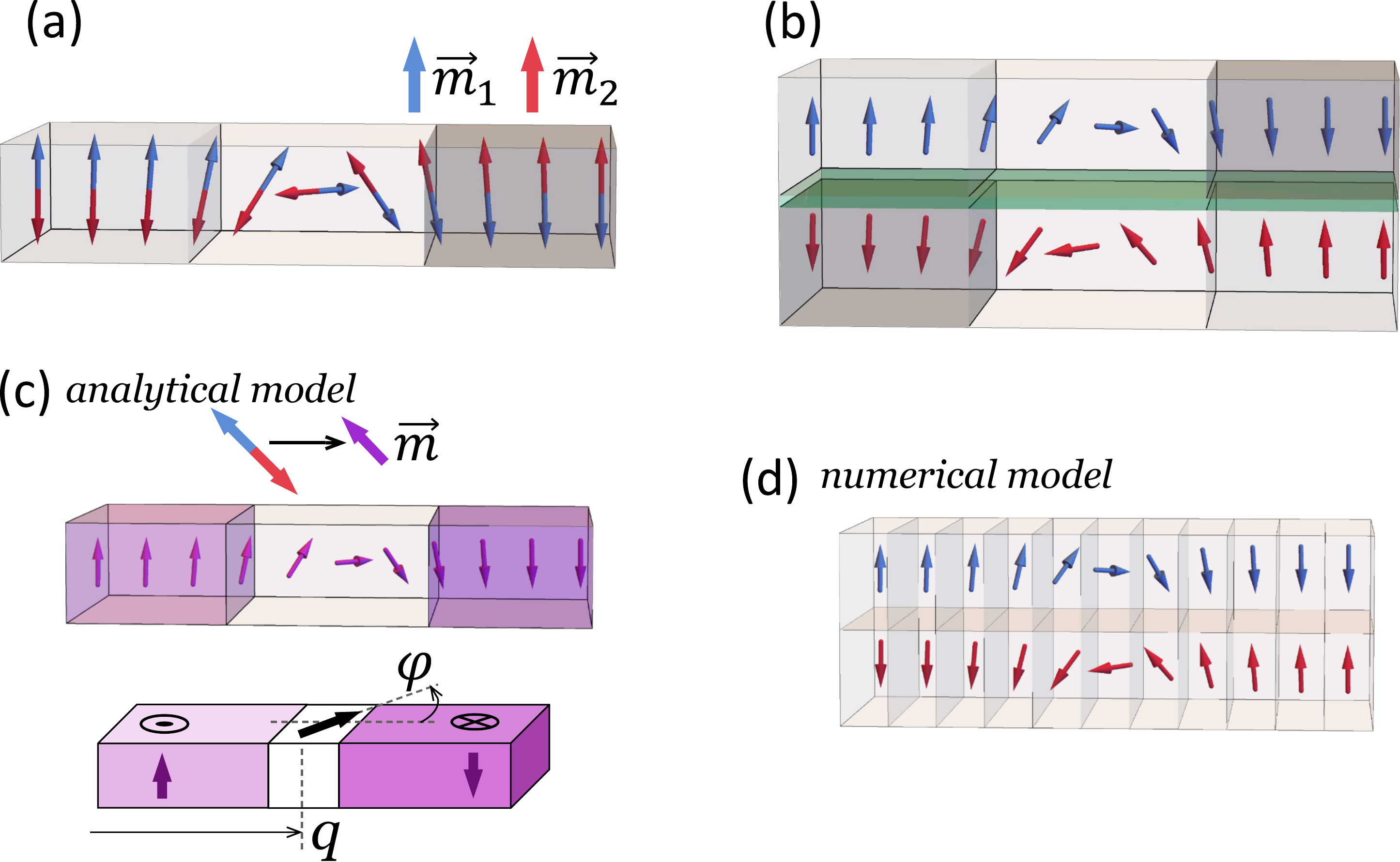}
	\caption{Illustrations of the two considered systems and of the used models.
	\textbf{(a)}~DW in a perpendicularly magnetized double-lattice system where the two lattices are spatially merged, such as in a ferrimagnetic alloy, and
	\textbf{(b)}~in a system with spatially-separated sublattices, such as a SAF.
	\textbf{(c)}~Analytical $q$-$\varphi$ model, with a single effective ferromagnetic DW with the effective parameters.
	\textbf{(d)}~Numerical, spatially-discretised model.
	}
	\label{fig:2DWferri}
\end{figure}

We use three models (two analytical and one numerical) of DW dynamics in double-lattice systems. The first is the well-studied ferromagnetic $q$-$\varphi$ analytical model.
We expand it to double-lattice systems using effective parameters~\cite{Hagedorn1972}, which defines an equivalent ferromagnetic material in the limit of infinite coupling between the sub-lattices (Fig.~\ref{fig:2DWferri}c). We will then compare it to the far less-restricted numerical micromagnetic double-lattice model (Fig.~\ref{fig:2DWferri}d). In section~\ref{sect:Low_Coupling}, where we study the effects of finite coupling and deviations from a perfect anti-alignment of the sublattices, we will use a double $q$-$\varphi$ analytical model (Fig.~\ref{fig:2DWferri_coupling}d).

\subsection{Effective parameters}

 The dynamics of a magnetisation distribution $\vec{M}(\vec{r},t)=M_S \vec{m}(\vec{r},t)$ is described by the Landau-Lifshitz-Gilbert equation (LLG)~\cite{Landau1935,Gilbert2004}:
 \begin{equation}
      \partial_t \vec{m} = -\mu_0\gamma  \vec{m}\times \oldvec{\mathbfcal{H}}  + \alpha  \vec{m}\times \partial_t \vec{m}+(\gamma/M_S)\vec{\tau}
      \label{eq_llg_old}
 \end{equation}
 $\gamma= M_S/L_S$ is the gyromagnetic ratio taken as positive in ferromagnets, $M_S$ is the spontaneous magnetisation, $L_S$ is the angular momentum density, $\alpha$ is the Gilbert damping parameter, $\oldvec{\mathbfcal{H}}\equiv\frac{-1}{\mu_0 M_S} \dmU$ is the effective field associated to $\dmU$ the variational derivative of the total energy density $U(\vec{m}(\vec{r}))$ (defined in sect.~\ref{sect:umagEnerg}). $\vec{\tau}$ accounts for non-conservative torques per unit volume applied on $\vec{m}$, such as STT or SOT.
  When dealing with effective parameters, it is helpful to express the LLG equation in terms of angular momentum density $L_S$ and of the product of the damping parameter $\alpha$ with $L_S$, $L_\alpha \equiv \alpha L_S$, which yields:
    \begin{equation}
     \partial_t \vec{m} = \frac{1}{L_S} \left( \vec{m}\times \dmU + L_\alpha \vec{m}\times\partial_t\vec{m} +\vec{\tau} \right)
    \label{eq:LLG_Full}
    \end{equation}
which can be rewritten in its explicit form:
    \begin{align}
    \nonumber
    \partial_t \vec{m}  = &  \frac{L_S}{L_S^2+L_\alpha^2} \left( \vec{m} \times \dmU + \vec{\tau}\right)  + \\
     +&\frac{L_\alpha}{L_S^2+L_\alpha^2} \vec{m} \times \left(  \vec{m} \times \dmU + \vec{\tau} \right)
    \label{eq:Solved_LLG}
    \end{align}
This form emphasises the two components of the dynamics of magnetisation. When $\vec{\tau}=0$, $\vec{m}$ precesses around $\dmU$ with rate $\left|\dmU\right|L_S/(L_S^2+L_\alpha^2)$ and relaxes towards $-\dmU$ with rate $\left|\dmU\right|L_\alpha/(L_S^2+L_\alpha^2)$.
The energy variation due to dissipation is \cite{Gilbert2004}
\footnote{
If $\vec{\tau}\neq 0$, the energy variation is instead
$d_tU= -L_\alpha \left| \partial_t \vec{m} \right|^2 + \frac{1}{L_S}\vec{\tau}\cdot \left( L_\alpha \partial_t \vec{m}+ \dmU   \right)$, or
$d_tU=-\frac{L_\alpha}{L_S^2+L_\alpha^2} \left( |\vec{m}\times\dmU|^2 -\vec{\tau}\cdot(\vec{m}\times\dmU) \right)+\frac{L_S}{L_S^2+L_\alpha^2} \dmU \cdot \vec{\tau}$
}
    \begin{equation}
    d_tU= -L_\alpha \left| \partial_t \vec{m} \right|^2 = -\frac{L_\alpha}{L_S^2+L_\alpha^2} \left|\vec{m}\times \dmU \right|^2
    \label{eq:Dissipation}
    \end{equation}
which is always negative (thanks to $L_\alpha >0 $) or zero when and only when $\vec{m}\parallel \dmU$, as it is physically expected.

The STT can be written as
$
    \vec{\tau}_{STT}=-(L_S\vec{u}\cdot\vec{\nabla})\vec{m}+ \vec{m} \times (\beta L_S \vec{u}\cdot\vec{\nabla})\vec{m}
$
with $L_S \vec{u} = P J \hbar/(2e) \vec{e}_J$, $\beta$ the non-adiabatic parameter~\cite{Thiaville2005}, $J \vec{e}_J$ the current density, and $P$ the current spin polarisation.
The SOT is separable in field-like (FL) and damping-like (DL) terms:
$
    \vec{\tau}_{SOT}=\tau_{\rm FL} \;\vec{m} \times \vec{\sigma}+\tau_{\rm DL}\;  \vec{m} \times \left( \vec{m} \times \vec{\sigma}\right)
$
with $\vec{\sigma}$ the orientation of the spin accumulation. When $\tau_{DL}$ is due to SHE, $\vec{\sigma}=\pm \vec{e}_J \times \vec{z}$ and $\tau_{\rm DL}=\tau_{\rm SHE}  = J \theta_{\rm{SHE}} \hbar/ (2 e t) $, with $t$ the film thickness and $ \theta_{\rm{SHE}}$ the SHE angle~\cite{Manchon2019}. \\

In the case of a multi-lattice material containing two magnetic populations (indexed `1' and `2'), the direction of the magnetisation of each sublattice ($\vec{m}_1$ or $\vec{m}_2$) can be described by the LLG equation with the respective sub-lattice parameters
(with individual spontaneous magnetisations $M_1$ and $M_2$, individual angular momentum densities $L_1$ and $L_2$, etc.)
with the total energy including now a inter-lattice coupling interaction $U_{12}=-J_{\rm AF}\vec{m}_1\cdot\vec{m}_2$ (with $J_{\rm AF}<0$ for antiferromagnetic coupling) \cite{Gomonay2014,Hrabec2018,Yang2015,Wangsness1953,Blasing2018}. In SAFs, RKKY provides a interfacial coupling term. Note that the dipolar interactions may add another inter-lattice interaction term. If the applied torques are much smaller than the inter-lattice coupling, the anti-alignment of magnetic moments $\vec{m}_1$ and $\vec{m}_2$ should not be significantly disturbed. If we consider the approximation $\vec{m}_1=-\vec{m}_2$, the magnetic order in the system can be described by an equivalent effective ferromagnetic film with $\vec{m}=\vec{m}_1=-\vec{m}_2$, similar as done in refs.~\cite{Wangsness1953,Hagedorn1972} and illustrated in Fig.~\ref{fig:2DWferri}c.
To find the parameters of the equivalent material, we consider the sum of the two individual LLG equations integrated along the thickness of the film, $(L_1 t_1) LLG_{1}(\vec{m})+ (L_2 t_2) LLG_{2}(-\vec{m})$,  which allows removing the coupling energy term. Then, the effective parameters can be deduced by comparing this sum to the effective LLG equation (also integrated along the thickness) $(L_S t) LLG (\vec{m} )$ term-by-term, yielding:
    \begin{eqnarray}
    \nonumber M_S &=&\left( M_1 t_1-M_2 t_2\right)/t\\
    \nonumber L_S &=&\left( L_1 t_1-L_2 t_2\right)/t\\
    L_{\alpha} &=& \left( L_{\alpha 1} t_1 + L_{\alpha 2} t_2 \right)/t
    \label{eq:Eff_Params}
    \end{eqnarray}

From hereon, the parameters of the sub-lattices are indexed 1 or 2, while the effective parameters of the equivalent ferromagnet are not. As before, the effective $\gamma$ is $M_S/L_S$ and the effective $\alpha$ is $L_\alpha/L_S$. If the two sub-lattices are in spatially separated layers with thicknesses $t_1$ and $t_2$, such as in a SAF or in a multilayer ferrimagnet, the effective thickness is $t=t_1+t_2$, while if the two lattices are spatially merged, such as in a single-layer ferrimagnetic alloy or antiferromagnet, $t=t_1=t_2$ instead. The MCP is reached when $M_1 t_1 = M_2 t_2$ and the ACP when $L_1 t_1 = L_2 t_2$. If the two sub-lattices are the same, $\gamma_1 = \gamma_2$ and the two compensation points coincide. Conversely, if the two sub-lattices are different (i.e. $\gamma_1 \neq \gamma_2$), the two compensation points are distinct~\cite{Hirata2018}.
A similar effective parameter approach could be applied to the case of ferromagnetic coupling or for more than two sub-lattices, with analogous equations. \\

 The effective $\dmU$ is determined using the derivation chain rule: $\dmU= \delta_{\vec{m}_1}U \partial_{\vec{m}} \vec{m}_1 + \delta_{\vec{m}_2}U  \partial_{\vec{m}}\vec{m}_2 = \delta_{\vec{m}_1}U-\delta_{\vec{m}_2}U$. The effective micromagnetic energy parameters (such as the exchange stiffness $A$, the uniaxial anisotropy $K$, or the DMI $D$, as defined in Appendix~\ref{sect:umagEnerg}) can then be deduced:
    \begin{eqnarray}
    \nonumber   A &=& (A_1 t_1+A_2 t_2)/t  \\
    \nonumber   K &=& \left(K_{1}t_1+K_{2}t_2\right)/t -K_{\rm dipolar} \\
                D &=& (D_1 t_1 +D_2 t_2)/t
    \label{eq:Eff_Params_microm}
    \end{eqnarray}
All subsequent system parameters, such as the DW width parameter $\Delta=\sqrt{A/K}$, can be calculated for the effective system by using the effective parameters. In the expression of $K$, we included $K_{\rm dipolar}$, the approximation of the dipolar-field-induced shape anisotropy valid in films thinner than the considered magnetic textures. The dipolar field is very different in antiferromagnetically-coupled systems with spatially merged and separated sub-lattices. With merged sub-lattices, $K_{\rm dipolar}=\frac{\mu_0}{2}M_S^2$ (zero at MCP as the dipolar field vanishes) and with spatially separated sub-lattices, $K_{\rm dipolar}=\frac{\mu_0}{2}(M_1^2 t_1 + M_2^2 t_2)/t$ (which is always finite). \\

Effective parameters can also be found for the current-induced torques:
    \begin{eqnarray}
    \nonumber P =& (P_{1} t_1-P_{2} t_2)/t \\
    \nonumber u   =&  \frac{\hbar/2}{L_S} \frac{P J}{e} \\
    \nonumber \beta =& (\beta_{1} P_1 t_1 + \beta_2 P_2 t_2)/(P t)\\
    \nonumber \tau_{FL} =&  (\tau_{FL 1} t_1 - \tau_{FL 2} t_2)/t \\
    \nonumber \tau_{DL} =&  (\tau_{DL 1} t_1 + \tau_{DL 2} t_2)/t \\
    \nonumber \theta_{\rm{SHE}} =&  \theta_{\rm{SHE},1} + \theta_{\rm{SHE},2} \\
    \tau_{\rm SHE}  =& J \theta_{\rm{SHE}} \hbar/ (2 e t)
    \label{eq:Eff_Params_J}
    \end{eqnarray}
Note that torques that are even in $\vec{m}$ add up (e.g. $\tau_{DL}$), while torques that are odd subtract (e.g. $\tau_{FL}$) \cite{Gomonay2014}. Different susceptibility to spin currents (STT or SOT) of each sub-lattice are accounted by the different polarization factors $P_1$ and $P_2$ and different spin Hall angles $\theta_{\rm{SHE},1}$ and $\theta_{\rm{SHE},2}$. \\

While in a ferromagnet almost all these quantities were constant and positive, the effective parameters, with the exception of $L_\alpha$, are signed and some may become zero or diverge ($\pm\infty$) at the compensation points (when $M_S\rightarrow 0$ or $L_S\rightarrow 0$) or when $P\rightarrow 0$.
All effective fields diverge at MCP (such as e.g. $H_{\rm DMI}=\frac{D/\Delta}{\mu_0 M_S}$ see Fig.~\ref{fig:pars}), and the parameters $\gamma$ and $\alpha$ diverge and change sign at ACP.  $\beta$ diverges at $P \rightarrow 0$. However, none of these divergences lead to non-physical behaviours (like e.g. an infinite $\partial_t \vec{m}$) as all the terms of  the LLG written in the form of  Eq.~\ref{eq:Solved_LLG} ($\dmU$, $L_S/(L_S^2+L_\alpha^2)$, $ \tau_{\rm SOT}$, $L_S u$, $L_S \beta u$, ...) remain non-infinite even at the compensation points. Likewise, even if $\alpha$ diverges at ACP, $L_\alpha$ remains finite and positive,  which ensures that the dissipation rate (eq.~\ref{eq:Dissipation}) is always negative.
In ref.~\cite{Kim2019}, a new definition of $\alpha$ was given to avoid divergence at ACP. However, this is not necessary as the dissipation rate (eq.~\ref{eq:Dissipation}) is not given by $\alpha$, which does diverge, but by $L_\alpha=\alpha M_S /\gamma$, which does not~\cite{Gilbert2004}.\\

Many effective parameters can be experimentally measured with usual techniques (e.g. $M_S$ and $K$ with magnetometry, current-induced torques by second harmonic Hall voltage~\cite{Krishnia2020}, ...), while sublattice parameters have to be selectively probed (e.g. x-ray dichroism in ferrimagnetic alloys).

\subsection{Collective coordinates model in the limit of strong coupling}

A very useful model of DW dynamics in perpendicularly-magnetised ferromagnetic systems is the 1D $q$-$\varphi$ equation~\cite{Schryer1974,Mougin2007,Thiaville2012} (Eq.~\ref{eq:qphi};  detailed in Appendix~\ref{sect:Appendix_qPhi}).
It describes the DW dynamics with two variables (the ``collective coordinates''): the DW position $q$ and its in-plane magnetisation angle $\varphi$ (see Fig~\ref{fig:2DWferri}).
Two steady-state DW propagation regimes are predicted~\cite{Schryer1974}: the translational regime at low drive, with a constant velocity and constant DW angle ($v=\dot{q}={\rm const.}$ and $\dot{\varphi}=0$), and the precessional regime at high drive, with oscillating velocity and precessing DW angle. The two regimes are separated by the Walker threshold~\cite{Schryer1974, Malozemoff1979, Mougin2007, Thiaville2012} (indicated by a subscript $W$)  that is determined by DMI and by the in-plane DW anisotropy created by the demagnetisation effect ($K_{\rm BN}$).
Note that for a DW driven by SHE alone, the precessional regime does not occur~\cite{Thiaville2012}. The $q$-$\varphi$ equation can be solved analyticall, and yields simple analytical laws for the translational regime and for the high-drive limit of the precessional regime (``asymptotic precessional regime'').We will also integrate it numerically for the full range of drive.
The analytical expressions for the velocity $v$, the internal angle $\varphi$ and the precession rate $\dot{\varphi}$ of a DW driven by field, SOT or STT, in  both regimes
are shown for reference in Table~\ref{tab:Classical_qphi}.

\begin{table*}[ht]
\centering
\begin{tabular*}{\textwidth}{l@{\extracolsep{\fill}}l|c|ll}
	\multicolumn{2}{l}{\textbf{Field-driven}} \\	\hline
	\multicolumn{2}{c|}{Translational regime}  & \multicolumn{1}{c|}{Walker threshold} &
	\multicolumn{2}{c}{Asympt. precessional regime} \\	\hline
	$v =\frac{\Delta}{L_\alpha} \mu_0 M_S H$ &
	$\varphi = \begin{cases}  \frac{1}{2} \arcsin\left(H/H_W \right) \\ \arcsin\left(H/H_W \right) \end{cases} $&
	$\left| H_W \right| = \begin{cases}
	                \left| \frac{L_\alpha}{L_S}\frac{1}{\mu_0 M_S} K_{\rm BN} \right| \\
	                \left| \frac{L_\alpha}{L_S}\frac{ 1}{\Delta\mu_0M_S} \frac{\pi}{2} D \right| \end{cases} $   &
	$v= \frac{\Delta L_\alpha}{L_S^2+L_\alpha^2} \mu_0 M_S H$ &
	$\dot{\varphi}= \frac{L_S}{L_S^2+L_\alpha^2} \mu_0 M_S H$ \\
	\hline
	\noalign{\vskip 3mm}
	\multicolumn{2}{l}{\textbf{STT-driven}} \\ \hline
	\multicolumn{2}{c|}{Translational regime}  & 	\multicolumn{1}{c|}{Walker threshold} &	\multicolumn{2}{c}{Asympt. precessional regime} \\
	\hline
	$v= \frac{\beta}{L_\alpha} L_S u$ &
	$\varphi=  \begin{cases}
	    \frac{1}{2} \arcsin(J/J_W) \\ \arcsin(J/J_W)
	 \end{cases}  $ &
	$\left|J_W \right| = \begin{cases}
	\left|\frac{L_\alpha}{L_S \beta P-L_\alpha P} \frac{e \Delta}{ \hbar /2} K_{BN} \right| \\
    \left|\frac{L_\alpha}{L_S \beta P-L_\alpha P} \frac{e}{ \hbar /2} \frac{\pi}{2} D \right| \end{cases} $ &
	$v = \frac{L_S+L_\alpha \beta}{L_S^2+L_\alpha^2} L_S u$ &
	$\dot{\varphi}= \frac{1}{\Delta}\frac{L_S \beta- L_\alpha}{L_S^2+L_\alpha^2} L_S u $\\
	\hline

	\noalign{\vskip 3mm}
	\multicolumn{2}{l}{\textbf{SOT-driven}} \\
	\hline
	$v= \frac{\Delta}{L_\alpha} \frac{\pi}{2}  \tau_{\rm SHE} \cos\varphi$ &
	$\varphi = \arctan \left(\tau_{\rm SHE}/ \frac{  L_\alpha D}{L_S \Delta} \right)$ & \multicolumn{1}{c|}{--}  & \multicolumn{2}{c}{--}\\
	\hline
\end{tabular*}
\caption[]{ $q-\varphi$ model solutions (from Table~\ref{tab:Classical_qphi}) with effective parameters for $v$, $\varphi$, $\dot{\varphi}$ and the Walker threshold of a DW driven by field, STT or SOT, in translational and asymptotic precessional regimes. The Walker thresholds have divergence points (discussed in the text), as do $\beta$ (at $P=0$) and $u$ (at $L_S=0$). However, $L_S u$, $\beta L_S u$ and $\beta P$ do not diverge. All other shown parameters do not diverge. The top and bottom expressions for $\varphi$ and Walker thresholds refer to the case of dominant dipolar-induced Bloch-Néel anisotropy ($K_{\rm BN}$) or dominant DMI ($D$). Here, $H_x=H_y=\tau_{\rm FL}=0$.
}
\label{tab:qphi_noDivergences}
\end{table*}

By applying the effective parameters described before to this model, it is possible to describe the DW dynamics in a multi-sublattice film, including across the compensation points. The resulting expressions for the DW dynamical parameters are shown in Table~\ref{tab:qphi_noDivergences} expressed with non-diverging parameters, wherever possible. Although some terms in eq.~\ref{eq:qphi} diverge at the compensation points, no unphysical behaviours are predicted by this model (e.g. infinite velocity), and all measurable quantities are finite.

\subsection{Numerical modeling of multi-lattice DW dynamics}

There may be deviations to the perfect anti-alignment of the two sublattices that is assumed in the model above. There could also be other distortions from the assumed 1D DW profile  To verify the validity of the analytical model, we have considered a numerical micromagnetic model by solving the LLG equations using the \textit{Mumax3} software~\cite{Vansteenkiste2014}. In section~\ref{sect:Low_Coupling} we will also address analytically the sublattice misalignment. To simulate a ferrimagnetic film, we use a simulation mesh with two layers, where the parameters of each layer correspond to the different sub-lattices of the film. An exchange coupling between mesh layers accounts for the antiferromagnetic inter-sub-lattice coupling.
We have not included the demagnetising field in the simulations of the CoGd, which is an acceptable approximation in the case of homogeneous and low magnetisation systems with a significant DMI~\cite{Thiaville2012}. The SAF was simulated in a similar way, but taking into account the demagnetising field, which is more important in this system, as well as the spatial gap between the layers. Due to the constraints of the spatial discretisation method, the simulations of the SAF were done for fewer values of $m_S$ than than those of the CoGd.

\section{DW dynamics \label{sect:DW_Dynamics}}

We examine now the DW dynamics under different driving forces (field, STT or SOT) using the models described above, in the two example systems: the ferrimagnetic alloy CoGd and a SAF CoFeB/Ru/CoFeB. We compare the analytical laws from the model (Table \ref{tab:qphi_noDivergences}; shown as lines in the plots) with the results of the numerical simulations (shown as points).
To facilitate the comparison between these two systems and other multi-sub-lattice systems, the plots will be expressed in terms of the normalised magnetisation  and normalised angular momentum density,
\begin{align}
    \nonumber m_S &\equiv (M_S t)/(|M_1 t_1|+|M_2 t_2|) \\
                l_S &\equiv (L_S t)/(|L_1 t_1|+|L_2 t_2|)
    \label{eq:msls}
\end{align}
The SAF will also be compared to a single layer system composed of only its layer 1 (i.e. $t_2=0$ and $m_S=l_S=1$).

\subsection{Material Parameters}
\begin{table}[ht]
\begin{tabular*}{8.5cm}{l@{\extracolsep{\fill}}|ll|ll}
                & \multicolumn{2}{c|}{\textbf{CoGd} }   & 	\multicolumn{2}{c}{\textbf{SAF} }   \\
                & \textbf{Co}     & \textbf{Gd}     & \textbf{`1'}     & \textbf{`2'}       \\ \hline
$M_s$ [MA/m]    & [0.82, 0.88]  & [0.89, 0.70]  & 0.8     & 0.8       \\ 
$t$ [nm]        & 3.8    & 3.8    & 0.8     & [0.6, 1.6] \\ 
$L_s$ [$10^{-6}$kg/ms]
                & [4.2, 4.5]  & [4.0, 5.1]  & 4.1     & 4.1 \\ 
Landé $g$       & 2.22    & 2.0   & 2.22    & 2.22      \\ 
$\alpha$        & 0.013  & 0.02   & 0.01   & 0.01     \\ 
$J_{\rm AF}$, $J_{\rm RKKY}$ & \multicolumn{2}{c|}{$-264$~MJ/m$^{3}$}  & \multicolumn{2}{c}{$-2$~mJ/m$^{2}$}   \\
$A$ [pJ/m]      & 16     & 0      & 16      & 16        \\ 
$K$ [kJ/m$^{3}$]& 300    & 0      & 450     & 450       \\ 
$D$ [mJ/m$^{2}$]& 0.1    & 0      & 0.8     & 0     \\ 
$P$             & 1      & 0      & 1       & 1         \\ 
$\beta$         & 0.6   & 0      & 0.1     & 0.1       \\ 
$\theta_{SHE}$  & 1      & 0      & 1       & 0       \\ \hline
$\Delta$ [nm]   & \multicolumn{2}{c|}{[7.3, 7.7]}  & \multicolumn{2}{c}{6}\\
$L_\alpha$ [nJs/m$^{3}$] &    \multicolumn{2}{c|}{[67, 80]}    &  \multicolumn{2}{c}{$41$} \\
\hline
\end{tabular*}
\caption{ Parameters used in the micromagnetic simulations and models, for the CoGd alloy (determined by mean field model for CoGd~\cite{Haltz2018}) and CoFeB/Ru/CoFeB SAF. Intervals (in brackets) refer to the range of variable quantities.
The RKKY coupling in the SAF is interfacial ($J_{\rm AF}\sim 1/t$), and so it is written as $J_{\rm RKKY}\equiv J_{\rm AF} t$. $P$ and $\theta_{\rm SHE}$ were taken as 1 for convenience.
}
\label{tab:pars}
\end{table}

\begin{figure*}[ht]
    \centering
    \includegraphics[width=\textwidth]{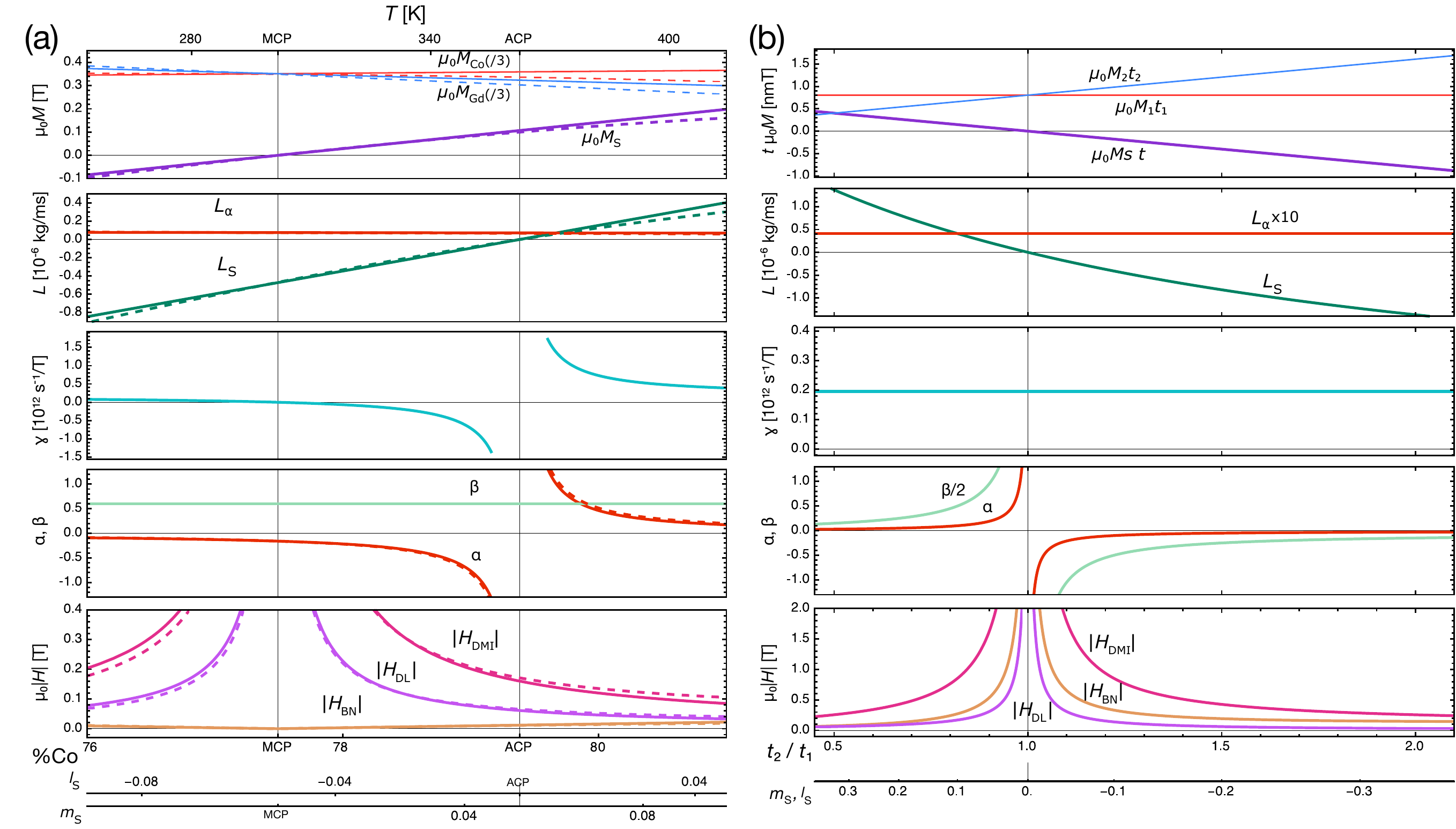}
    \caption{
    \textbf{(a)} Effective parameters of the CoGd ferrimagnetic alloy versus temperature and composition (solid and dashed lines, respectively), based on a mean-field model:
    $M_S$ (and sublattice magnetisation), $L_S$, $L_\alpha$, $\gamma$, $\alpha$, $\beta$, and the effective fields $|H_{\rm BN}|$, $|H_{\rm DMI}|$
    , $|H_{\rm DL}|=|\tau_{\rm DL}/(\mu_0 M_S)|$ (at $J=30$~GA/m$^2$).
    \textbf{(b)} Effective parameters for the SAF system, versus layer thickness ratio. $l_S$ and $m_S$ are shown in the horizontal bars on the bottom.}
    \label{fig:pars}
\end{figure*}

Table~\ref{tab:pars} shows the parameters used for these two systems. The chosen parameters are close to experimentally reported values when these were available \cite{Yang2015,Hansen1989,Krishnia2020}; some were chosen for the convenience of presenting the DW dynamics (e.g. $P=\theta_{\rm SHE}=1$ and constant $K$).
For CoGd, the variation with temperature and with alloy composition of the sub-lattice magnetisations, $M_{\rm Co}$ and $M_{\rm Gd}$, were calculated using mean field theory and the respective angular momenta were calculated using the Landé $g$ factors in the table (see refs.~\cite{Haltz2018,Hansen1989} for details). Fig.~\ref{fig:pars}a shows the variation of magnetisation and some effective parameters of the CoGd film versus alloy composition at a fixed temperature (300~K) (full lines and bottom axis) or versus temperature at fixed composition (\%Co=75\%) (doted curves and top axis). The variations with temperature and with composition superpose well, which means that the same set of parameters can be obtained equally by changing the composition or the temperature, in the considered ranges. The two compensation points, MCP and ACP, are clearly visible. Since the two sub-lattices are of different nature ($\gamma_1\neq \gamma_2$) the two compensation points are distinct. $\gamma$ changes sign twice (crossing zero at the MCP and diverging at the ACP) and $\alpha$ once (diverging at ACP). In the CoGd film, we consider current-induced torques only in the Co sub-lattice ($P_{\rm Gd}=\theta_{\rm SHE, Gd}=0$), which is a common assumption in RE-TM alloys where mostly the TM sub-lattice is active in the spin transport~\cite{Haltz2019}.

To tune the sub-lattices in the SAF, we considered a layer `1' with a fixed thickness ($t_1=0.8$~nm) and layer `2' with variable thickness. Fig.~\ref{fig:pars}b shows the variation of the effective parameters of the SAF versus the layer thickness ratio $t_2/t_1$.
In contrast with CoGd, the two sub-lattices are of the same material (with $\gamma_1 = \gamma_2$) and so the two compensation points coincide (at $t_2/t_1=1$). $\alpha$ diverges and changes sign at the compensation point but $\gamma$ is always non-zero and finite. For this system, the DMI and the SHE torque are assumed to be present only in layer `1', which corresponds to a SAF structure with a single heavy metal adjacent layer. The STT is applied in the two layers with the same magnitude ($P_1=P_2$ and $\beta_1=\beta_2$).\\

\subsection{DW driven by field}

\begin{figure*}[ht]
	\centering
	\includegraphics[width=\textwidth]{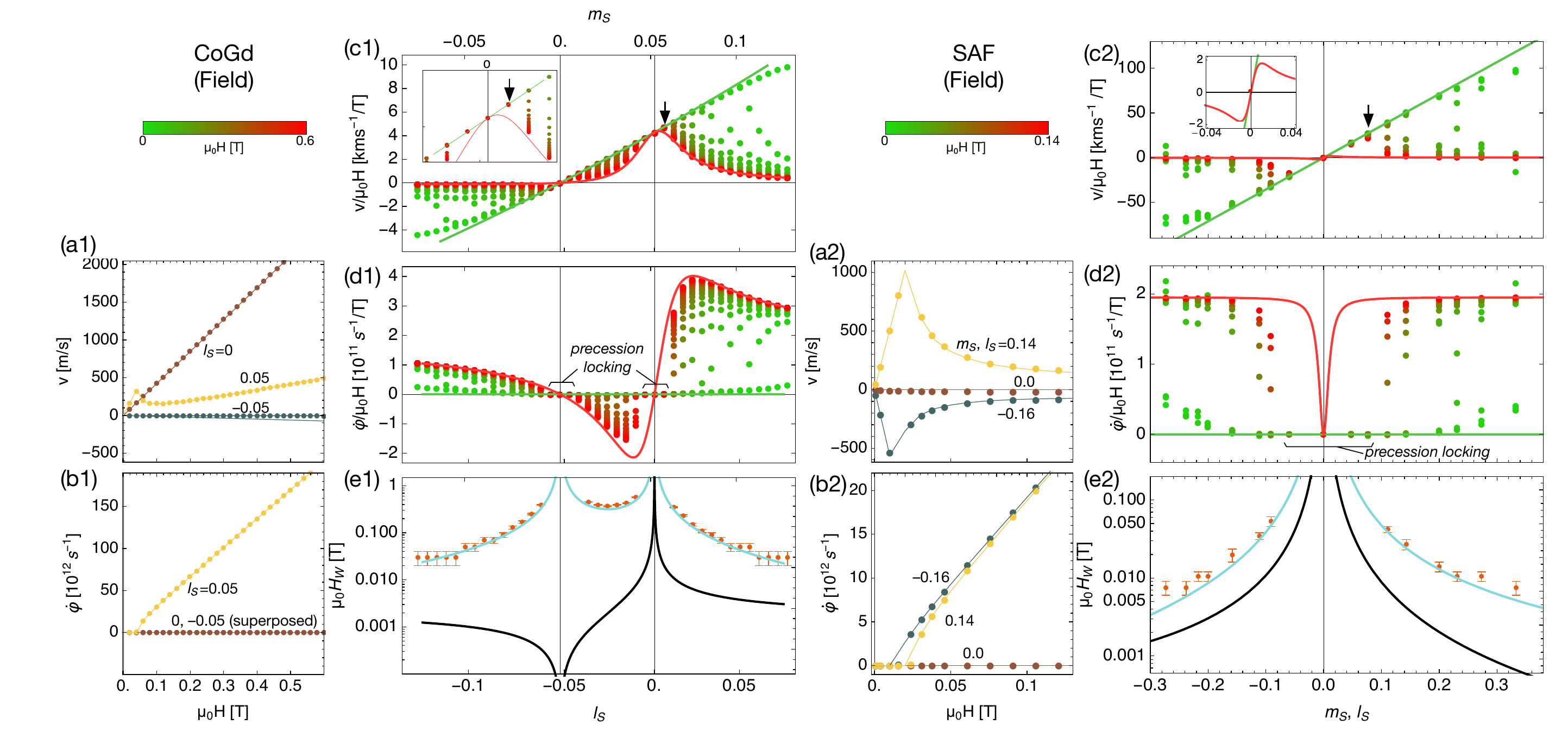}
    \caption{
    Field-driven DW dynamics in CoGd (a1-e1) and in the SAF (a2-e2).
    \textbf{(a1,2)} Velocity $v$ and
    \textbf{(b1,2)} precession rate $\dot{\varphi}$ versus $H$ for a few values of $l_S$. Points correspond to simulations and lines to the $q$-$\varphi$ model (eq.~\ref{eq:qphi}).
    \textbf{(c1,2)} Mobility $v/\mu_0 H$ and
    \textbf{(d1,2)} normalised precession rate $\dot{\varphi}/\mu_0 H$ versus $l_S$. The points correspond to micromagnetic simulations (low fields in green and high fields in red; see colour scale), and the lines to the translational (green) and asymptotically precessional (red) regimes of the model (Tab.~\ref{tab:qphi_noDivergences}).  The insets in (c1,2) show the region close to $l_S=0$. The arrows indicate the point of maximum mobility at the highest field.
    \textbf{(e1,2)} Walker field $H_W$ versus $l_S$. Points are extracted from (d1,2) and the lines correspond to the model for DMI dominated (blue) and dipolar-field dominated (black) cases.
    }
	\label{fig:DWM_H}
\end{figure*}

Fig.~\ref{fig:DWM_H} a1,2 and b1,2 show the DW velocity $v$ and precession rate $\dot{\varphi}$ versus the perpendicular field $H$ in CoGd and in the SAF, for different values of $l_S$. The numerical integration of the $q$-$\varphi$ equation (eq.~\ref{eq:qphi}; lines) matches very well the numerical simulations (points).
Fast DW velocities are obtained (a few km/s), which vary in magnitude and direction with $m_S$, similar to what is reported in experiments~\cite{Kim2017b}. The translational ($\dot{\varphi}=0$) and precessional ($\dot{\varphi}\neq0$) regimes are clearly visible,  but do not occur for every $l_S$ (e.g. for $m_S=0$ or $l_S=0$).

The dynamics of the DW versus $l_S$ is more clearly seen in Fig.~\ref{fig:DWM_H}c1,2 and d1,2 where we show the mobility ($v/\mu_0 H$) and the normalised precession rate ($\dot{\varphi}/\mu_0 H$) versus $l_S$ for CoGd and for the SAF.  Except close to ACP, the simulated DW (points) follows the analytical law for the translational regime (Tab.~\ref{tab:qphi_noDivergences}; green line) at low fields and approaches at large fields the law of the asymptotically precessional regime (red line). In general, the SAF shows significantly higher mobility than CoGd, as the mobility is inversely proportional to  $L_\alpha$ (Tab.~\ref{tab:qphi_noDivergences}), which is much lower in the SAF (Fig.~\ref{fig:pars}).
For both CoGd and SAF, the mobility is reversed and crosses zero at MCP. Physically, this can be understood since the effect of a (moderate) external field is reversed when the sub-lattice with the largest magnetisation changes (i.e. $M_S$ is reversed) and produces no effect if the sub-lattices are compensated ($M_S=0$). In the \textit{q}-$\varphi$ model (Table \ref{tab:qphi_noDivergences}), this is reflected in both regimes as $v\propto M_S H$.
The mobility of the precessional regime is comparable to that of the translational regime in CoGd, while in the SAF it is much smaller. The ratio of precessional to translational mobility is given by $(1+L_S^2/L_\alpha^2)^{-1}$ (Table~\ref{tab:qphi_noDivergences}), which is indeed much smaller in the SAF than in the CoGd (compare Figs.~\ref{fig:pars}a and b).

The \textit{q}-$\varphi$ model also predicts that the DW mobility in the translational regime is not directly dependent of $L_S$, and so it is not affected by the ACP. In contrast, the mobility in the precessional regime is maximum at ACP as $v/H\propto  M_S(L_S^2+L_\alpha^2)^{-1}$. In the SAF, as the ACP and MCP coincide, the precessional regime mobility shows two maxima (of opposite sign) on either side of the compensation point and is zero at that point (see inset of Fig.~\ref{fig:DWM_H}c2). \\

In CoGd, $\dot{\varphi}$ crosses zero twice at MCP and ACP. Particularly, at the ACP, the DW propagates without any precession ($\dot{\varphi}=0$).
The \textit{q}-$\varphi$ model predicts that indeed $\dot{\varphi} \propto L_S M_S H$ crossing zero at MCP and ACP and reversing its direction between these two points. This is closely related to the magnetic precession rate under field given by the LLG equation (eq.~\ref{eq:Solved_LLG}), which is $\propto \dmU L_S = \mu_0 H M_S L_S$.
For the SAF, as the two compensation points are superimposed, there is no intermediate region of reversed precession and $\dot{\varphi}$ is zero at compensation when both $L_S=0$ and $M_S=0$. At the point of no precession, the velocity of the translational and precessional regimes predicted by \textit{q}-$\varphi$ coincide (see eq.~\ref{eq:qphi}). \\

In the simulations, the precession rate is zero in the vicinity of the compensation point for all tested fields (see Fig.~\ref{fig:DWM_H}d1,2 and inset of c1). We call this case precession locking. This phenomenon occurs generally around every point of no precession, and can be understood by analysing the Walker field $H_W$, shown in Fig.~\ref{fig:DWM_H} e1 and e2 for the CoGd and SAF, respectively.
The \textit{q}-$\varphi$ model solutions are shown with DMI (blue line) and with $K_{\rm BN}$ and no DMI (black line; calculated as described in Appendix~\ref{sect:Appendix_qPhi}).  We note that the SAF shows much higher $H_W$ than the single layer film of the same material (0.85~mT). The $H_W$ from the simulations follow the DMI solution, as expected, and diverge at ACP and MCP. In the vicinity of the $H_W$ divergence, the DW will propagate in the translational regime even for the highest applied field --
leading to precession locking.  The model predicts indeed that for the case of dominant DMI, $H_W\propto D L_S^{-1} M_S^{-1}$. Without DMI, $H_W\propto K_{\rm BN} L_S^{-1} M_S^{-1}$.  For homogeneous ferrimagnets like CoGd, $K_{\rm BN} \propto M_S^2$, and so $H_W$ diverges at ACP but not at MCP (and is quite small far from ACP). For separated systems like the SAF, $K_{\rm BN}$ remains finite at the compensation, and thus $H_{\rm W}$ diverges with or without DMI.

 An interesting consequence of precession locking is that the maximum mobility at a given applied field does not occur exactly at ACP. If the DW dynamics versus temperature is measured in CoGd at a constant high field, two points of discontinuity will be observed close to ACP corresponding to $H=H_W$. At these points, the DW switches between propagation regimes and its velocity and precession rate change abruptly.  This transition can be seen in Fig.~\ref{fig:DWM_H}c1 where the points at highest field (in red) follow the precessional curve (in red) but switch to the translational curve (in green) near ACP.  As a result, the maximum of mobility at high field is not exactly at ACP (see arrows in Fig.~\ref{fig:DWM_H}c1). As the mobility of the two propagation regimes converge at ACP, these discontinuities may go unnoticed experimentally, but they may induce errors in the determination of the ACP from the DW mobility peak \cite{Haltz2020}. The same phenomenon should be observable in a SAF by measuring velocity versus layer thickness (for example by using a wedge layer). \\

The mobility peak near ACP was already experimentally observed in a ferrimagnetic alloy (GdFeCo) by Kim et al.~\cite{Kim2017b}. The authors observed that the DW mobility increased as the system was heated to a temperature above MCP, up to $\sim 20$~kms$^{-1}$/T at $\sim90$~K above MCP, and then decreased quickly after. The authors identified the maximum with ACP, by assuming that the DW was in the precessional regime at the peak, although these models predict that it was in the translational regime, and that the maximum is located slightly above ACP.

\subsection{DW driven by STT}

\begin{figure*}[ht]
\centering
\includegraphics[width=\textwidth]{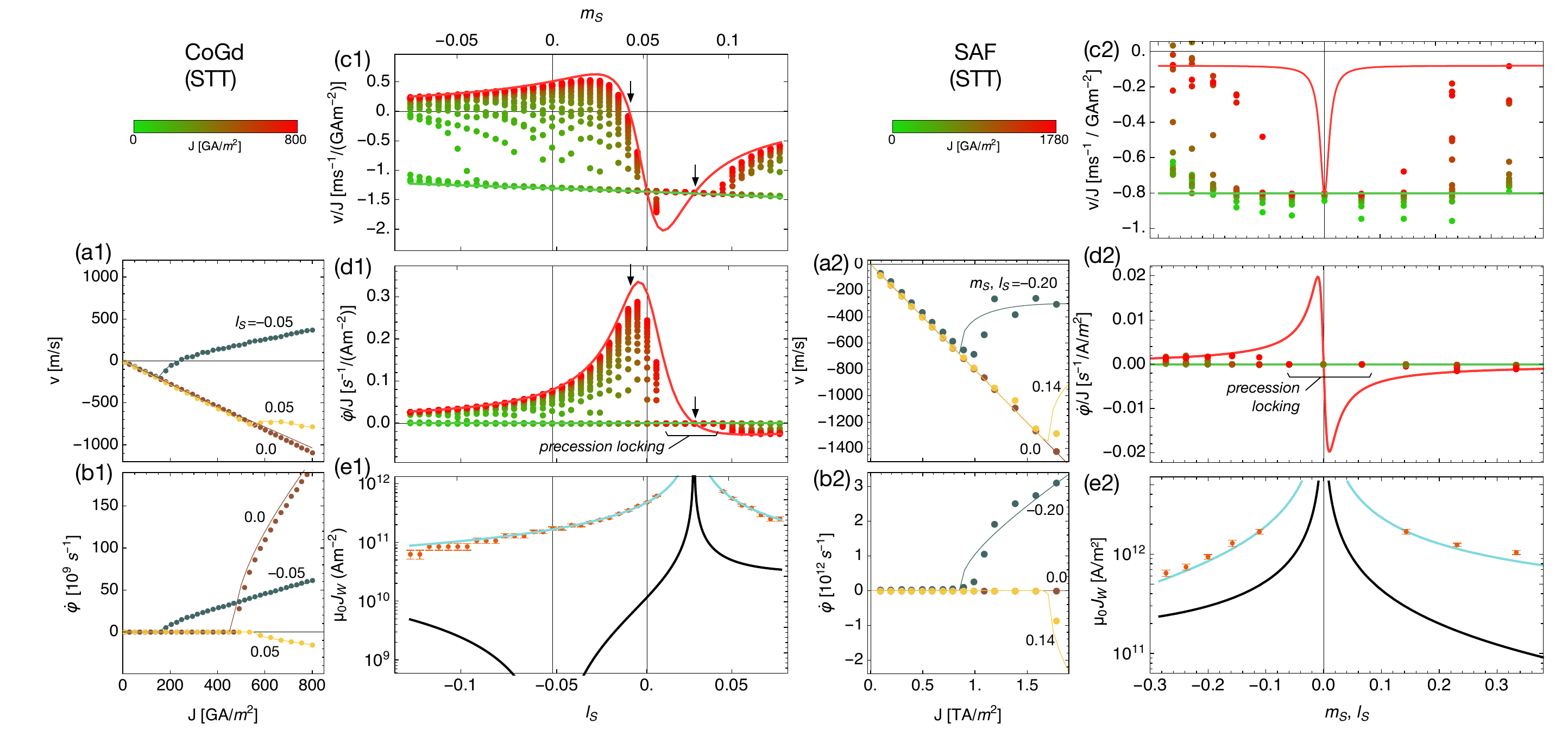}
    \caption{
    STT-driven DW dynamics in CoGd (a1-e1) and in the SAF (a2-e2).
    \textbf{(a1,2)} Velocity $v$ and
    \textbf{(b1,2)} precession rate $\dot{\varphi}$ versus $J$ for a few values of $l_S$. Points correspond to numerical simulations and lines to the $q$-$\varphi$ model (eq.~\ref{eq:qphi}).
    \textbf{(c1,c2)} Mobility $v/J$ and
    \textbf{(d1,d2)} normalised precession rate $\dot{\varphi}/J$ versus $l_S$. The dots correspond to micromagnetic simulations (low currents in green and high currents in red; see colour scale). Lines correspond to the translational (green) and asymptotically precessional (red) regimes of the model (Tab.~\ref{tab:qphi_noDivergences}). The arrows indicate the point of zero mobility and zero precession, discussed in the main text.
    \textbf{(e1,2) }Walker current $J_W$ versus $l_S$. Dots are extracted from (d1,2) and the lines correspond to  $q$-$\varphi$ model for DMI dominated (blue) and dipolar-field dominated (black) regimes.}
\label{fig:DWM_STT}
\end{figure*}

Under STT, two non-conservative torques act on the DW, inducing a more complex behaviour than with field. We consider that the current interacts only with the Co sub-lattice of CoGd, whereas it interacts with both layers of the SAF (see Table~\ref{tab:pars}). Figs.~\ref{fig:DWM_STT}a1,2 and b1,2 show the DW velocity $v$ and precession rate $\dot{\varphi}$ versus $J$ for CoGd and the SAF with different values of $l_S$. The agreement between simulations (points) and the integrated \textit{q}-$\varphi$ model (lines) is very good.
As in ferromagnets and in the field-driven case, two propagation regimes are visible:  the translational regime at low current ($\dot{\varphi}=0$) and the precessional regime at high current ($\dot{\varphi}\neq0$). Interestingly, in the precessional regime in CoGd, the propagation and precession directions change with $l_S$ and with $J$.
\\

The variation with $l_S$ is more visible in Fig.~\ref{fig:DWM_STT}c1,2 and d1,2 that shows the mobility ($v/ J$) and the normalised precession rate ($\dot{\varphi}/ J$) versus $l_S$ and $m_S$.
For low current (green points), the simulated DW follows the translational laws (green lines) and, for higher current (red points), the DW approaches the laws for the asymptotic precessional regime (red lines). In CoGd (for which MCP$\neq$ACP), there are no remarkable variations of these two quantities near MCP, showing that the STT does not depend directly on $M_S$.

In both systems, the mobility of the translational regime does not vary significantly with $l_S$ when compared with the precessional regime. Indeed, according to the model, the DW mobility in the translational regime is $\beta L_S u/L_{\alpha}$ (Table~\ref{tab:qphi_noDivergences}) which is almost constant in the investigated range of parameters (as $\beta L_S u$ is almost constant, eq.~\ref{eq:Eff_Params_J}). Note that the mobility does not depend on $M_S$.
Indeed, although the STT velocity in ferromagnets could be thought to be $\propto u \propto 1/M_S$ (Tab.~\ref{tab:Classical_qphi}), once it is explicitly written with non-diverging effective parameters (Tab.~\ref{tab:qphi_noDivergences}) no direct dependence on $M_S$ exists.

In the precessional regime the DW behaviour is more complex. In CoGd, there are two remarkable points (marked with arrows in fig.~\ref{fig:DWM_STT}d1,c1). Slightly before ACP, the propagation direction reverses and the precession rate exhibits a peak. Slightly after ACP, the precession rate changes its sign. In the SAF, no propagation reversal occurs, but the mobility is maximum near ACP.

The \textit{q}-$\varphi$ model predicts that the reversal of $v$ ($v=0$) in the precessional regime occurs at $L_S P=-\beta L_\alpha P$. For CoGd, $\beta=\beta_1>0$, so the reversal occurs just before ACP. For the SAF, in the studied range, this condition is never satisfied and no reversal occurs \footnote{If $P_1$ and $P_2$ have the same sign, $v=0$ occurs at $L_S=-\beta L_\alpha$. If the signs are opposite, it occurs at $\beta_1 P_1 t_1+\beta_2 P_2 t_2=0$ since $\beta_1$ and $\beta_2$ are positive.}.

The precession-free point ($\dot{\varphi}=0$) is predicted at $L_S \beta P= L_\alpha P$. In CoGd, it does not occur at ACP but above, at $L_S= L_\alpha/\beta$ (equivalently, $\beta=\alpha$, visible in Fig.~\ref{fig:pars}a). For the SAF it occurs at compensation (where $L_S=P=0$), although for a SAF with different materials this point would occur elsewhere as $P\neq0$ at compensation.

 As with field-driven DWs, in the simulations the precession is locked (zero) in the vicinity of the no-precession point predicted by the $q$-$\varphi$ model. This can be understood by examining the Walker threshold current $J_W$, show in Fig.~\ref{fig:DWM_STT}c1,2 for CoGd and the SAF. The $J_W$ extracted from the simulations (points) follows closely the prediction for the DMI dominant case (Tab.~\ref{tab:qphi_noDivergences}; blue line).  The SAF shows a much higher $J_W$ than a single layer of the same material ($4 \times 10^{11}$~A/m$^2$). With DMI, $J_W>0$ everywhere. Without DMI (black line), the $J_W$ is proportional to $K_{BN}$, which is zero at MCP for systems with spatially-merged sub-lattices (CoGd) but is always finite in SAFs.
For both cases, the $q$-$\varphi$ model predicts that the Walker threshold $J_W$ diverges when $L_S\beta P = L_\alpha P$, which corresponds to the previously-discussed precession-free point. For a given maximum $J$, in the vicinity of this point the Walker current cannot be reached and the DW is precession-locked in the translational regime. \\

Some of these aspects of the STT-driven DW dynamics have recently been observed by Okuno et al.~\cite{Okuno2019} in ferrimagnetic GdFeCo. The authors measured the DW velocity driven simultaneously by field and current for various temperatures. By assuming a linearity between field and current, the authors separated the mean velocity due to the field and the difference of velocities due to the STT. They observed a peak of the STT-induced velocity versus temperature that they attributed to the ACP (with a limited validity, as discussed above) and a reversal of the STT-induced velocity close to this temperature.\\

\subsection{DW driven by SHE}

\begin{figure*}[ht]
    \centering
    \includegraphics[width=\textwidth]{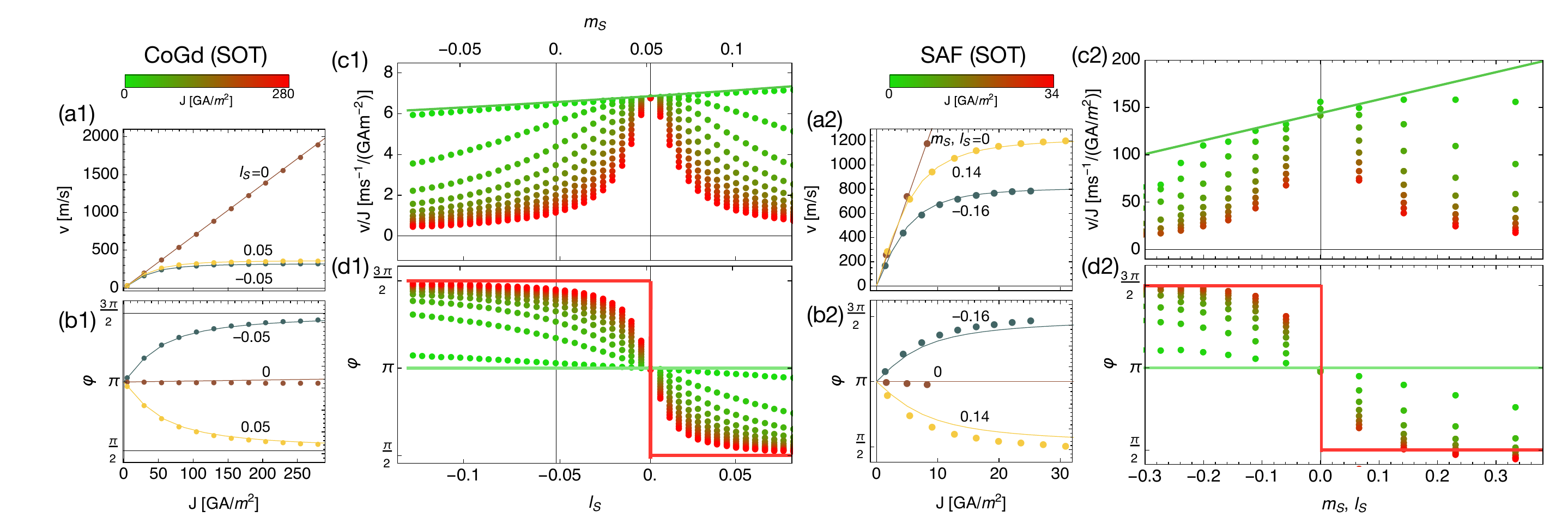}
    \caption{
    SHE-driven DW dynamics in CoGd (a1-d1) and in the SAF (a2-d2).
    \textbf{(a1,2)} Velocity $v$ and
    \textbf{(b1,2)} in-plane magnetisation angle $\varphi$ versus $J$ for different values of $l_S$. Points correspond to numerical simulations and lines to the $q$-$\varphi$ model (eq.~\ref{eq:qphi}).
    \textbf{(c1,2)} Mobility $v/J$ and
    \textbf{(d1,2)} in-plane angle $\varphi$ versus $l_S$. The dots correspond to micromagnetic simulations (low currents in green and high currents in red; see the color scale), lines to the asymptotes of the $q$-$\varphi$ model (Tab.~\ref{tab:qphi_noDivergences}) at low (green) and high current (red). At high current, the model predicts that the velocity is asymptotically limited and the mobility tends to zero, except at ACP.
    }
    \label{fig:DWM_SOT}
\end{figure*}

We describe here the DW driven by SHE-induced SOT, although the results also apply to any other sources of damping-like torque of the same form.
We do not consider other current-induced torques like the $\tau_{\rm FL}$, which are typically much smaller and lead to more complex dynamics (similar to the STT-driven case). As discussed previously (Table~\ref{tab:pars}), the SOT is present only in the Co sublattice of CoGd and in the layer `1' of the SAF.

Figs.~\ref{fig:DWM_SOT}a1,2 and b1,2 show the DW velocity $v$ and the DW in-plane angle $\varphi$ versus current density $J$ for CoGd and SAF, with different values of $l_S$.
The numerical integration of the $q$-$\varphi$ equation (lines) show an excellent agreement with the micromagnetic simulations (points).
As in ferromagnets, the SOT-driven DW only shows a translational regime ($\dot{\varphi}=0$). $\varphi$, initially $\approx 0$ at low current, approaches $\pm\pi/2$ with increasing current (as predicted by the model: $\varphi={\rm arctan}(\tau_{\rm SHE} \frac{\Delta L_S}{D L_\alpha})$). As $v\propto \cos\varphi$, the velocity saturates at high current ($v\rightarrow\frac{\pi}{2}D/L_S$). While the propagation direction is always the same, the saturation velocity and $\varphi$ strongly vary with $l_S$. In particular, at ACP, the behaviour is remarkably different from the ferromagnetic case: the DW propagates without magnetisation tilt ($\varphi$ remains 0) and no saturation is visible.

Fig.~\ref{fig:DWM_SOT} c1,2 and d1,2 shows the plots of mobility ($v/J$) and $\varphi$ versus $l_S$. The lines are the limits at low (green) and high (red) current given by the model, and the points correspond to the simulations (green for low current, red for high current). For the CoGd, no remarkable features appear near MCP, showing that the SHE drive, as the STT, does not depend directly on $M_S$. This contrasts with the effective field created by the DL torque ($H_{\rm DL}$, shown in Fig.~\ref{fig:pars}), which diverges at MCP, showing that the effective field is not a convenient parameter to characterise the effects of SOTs in these systems.
At low current, $\varphi$ remains $\approx0$ for all values of $m_S$ and the mobility is large and almost constant (as predicted: $v/\tau_{\rm SHE}\approx\frac{\pi}{2}\Delta /L_\alpha$). At higher current, both quantities vary strongly with $m_S$. In particular, at ACP, $\varphi=0$ and changes sign while the mobility shows a peak independent of $J$.
The model indeed predicts that the velocity does not saturate at ACP (the saturation velocity, $\frac{\pi}{2}D/L_S$, diverges), which implies a mobility peak for all currents.
For the SAF, this means that the DW can be driven much faster than in the single layer case (i.e. $t_2=0$, which presents a saturation velocity of 306~m/s).
\\

Several experimental reports have shown some of these effects. The variation of the SOT-driven DW velocity with the thickness ratio of a SAF was observed by Yang et al.~\cite{Yang2015}.  The authors measured the DW dynamics in perpendicular-magnetised SAFs with a single Pt adjacent layer, and various layers thicknesses. Both the observed DW mobility and the maximum velocity were greatest at compensation, when the two constituting layers of the SAF were balanced.

Recently, several experimental reports have studied SOT-driven DW dynamics in ferrimagnetic systems.  Siddiqui et al.~\cite{Siddiqui2018} measured the SHE-driven DW velocity in Pt/CoTb/SiN tracks with variable composition observing, for the first time, a mobility peak (versus composition) that could be associated to the ACP. However, the investigated velocity range and the threshold current (required to depin the DW) make it difficult to unambiguously attribute this mobility peak to the ACP. Also, different compositions were measured in different structures, which can present deviations of parameters such as the anisotropy or the DW pinning field.

Caretta \textit{et al.}~\cite{Caretta2018} studied the SHE-driven DW in Pt/CoGd/TaOx tracks with varying temperature. These experimental results are in very good agreement with the presented model. At high current, a velocity peak versus temperature was observed, which was attributed to the ACP. At low current, however, the peak fades and the velocity did not depend significantly on the temperature.

In our recent work, we studied the SHE-driven DWs in CoFeGd/Pt with variable temperature~\cite{Haltz2020}. A transverse bias field was applied and, using an extension of the described $q$-$\varphi$ model to include bias fields, the $\varphi$ could be determined. A mobility peak (versus temperature) was observed, at which $\varphi$ vanished and changed sign, showing the precession-free dynamics of the DW at ACP.

Similar conclusions on the velocity versus $l_S$ in ferrimagnets were recently obtained from a theoretical point of view using micromagnetic models and an extended 1D model in \cite{Martinez2019}.

\section{Effects of finite coupling \label{sect:Low_Coupling}}

It is important to investigate how the results presented in the previous sections are modified by a finite angle between the sub-lattice magnetisations ($\vec{m}_1\neq-\vec{m}_2$). The strength of the inter-lattice coupling in ferrimagnetic alloys varies significantly for different alloy compositions or deposition techniques \cite{Gambino1978,Hebler2016,Yang2019}, and whether the RE-TM is an alloy or is spatially-segregated in multilayers. In SAFs, the coupling strength is much weaker than in RE-TM alloys, and can be modulated (even made ferromagnetic) by altering the spacer layer~\cite{Parkin1990}.

\subsection{Collective coordinates model for finite coupling \label{sect:Model_Collective_Finite}}

\begin{figure}[ht]
	\centering
	\includegraphics[width=\columnwidth]{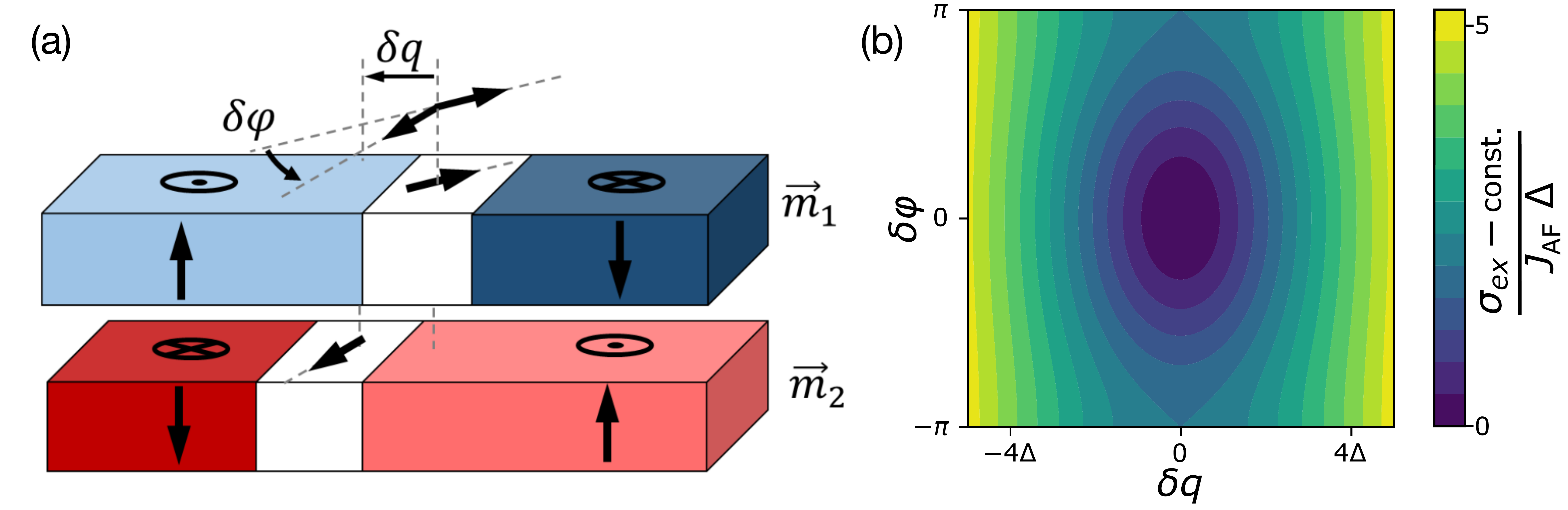}
	\caption{
	\textbf{(a)} Diagram of $\delta q$ and $\delta \varphi$.
	\textbf{(b)} Surface energy plot of $\sigma_{ex}$ versus $\delta q$ and $\delta\varphi$.}
	\label{fig:2DWferri_coupling}
\end{figure}

To study the effects of finite coupling, we developed a collective coordinates model with two coupled sub-lattices. We consider a DW in a double-lattice system as two rigid ferromagnetic DWs, one for each sub-lattice, coupled to each other~\cite{Blasing2018,Yang2015} (Fig.~\ref{fig:2DWferri_coupling}a).  The choice of this ansatz for the DW profile is motivated by the fact it is exact in the two limiting cases of infinite and null interlattice coupling.
We consider an $\uparrow\downarrow$ DW in the sublattice 1, and a $\downarrow\uparrow$ DW in the sublattice 2, with coordinates $(q_1,\varphi_1)$ and $(q_2,\varphi_2)$, respectively.
The sub-lattice 1 is arbitrarily chosen as the main one, and we define four more convenient coordinates $q$, $\delta q$, $\varphi$ and $\delta\varphi$ (shown in Fig.~\ref{fig:2DWferri_coupling}a), such as:
    \begin{align*}
    q_1 &=q,            & \varphi_1 &= \varphi \\
    q_2 &=q +\delta q,  & \varphi_2 &= \varphi+ \pi+ \delta\varphi
    \end{align*}
This model allows for some inter-lattice tilt in the form of lag between the positions of the DWs ($\delta q$) or an angle difference between their central antiparallel magnetisations ($\delta \varphi$).  A similar model was considered in refs.~\cite{Yang2015,Martinez2019}, although without the $\delta q$ degree of freedom.

The time evolution of the four collective coordinates $q$, $\varphi$, $\delta q$, $\delta \varphi$ are described by using two coupled $q-\varphi$ equations with different parameters for each sublattice, with an additional energy term, $\sigma_{\rm ex}$, that accounts for the interlattice coupling  energy per unit of DW surface. (Appendix~\ref{sect:Appendix_qPhi}).
We estimate $\sigma_{\rm ex}$ from the classical DW profile, $\sigma_{\rm ex} = \int_{-\infty}^{+\infty} U_{12}(x) dx = \int_{-\infty}^{+\infty} -J_{AF}\vec{m}_1(x) \cdot \vec{m}_2(x) dx$, which yields:
\begin{align}
\nonumber  \sigma_{\rm ex} & =
-2 J_{\rm AF} \Delta  \frac{ \delta q/\Delta}{{\rm sinh}  \frac{\delta q}{\Delta}}\left( \cosh\frac{\delta q}{\Delta}-\cos\left(\delta\varphi\right) \right) + {\rm const.} \\
    & \approx  -J_{\rm AF} \Delta \left(\delta\varphi^2+\left(\delta q/\Delta \right)^2 \right) + {\rm const.}
\label{eq:Sigmaex}
\end{align}
where the approximation is valid for small $\delta q/\Delta$ and $\delta \varphi$. Fig.~\ref{fig:2DWferri_coupling}b shows the inter-exchange coupling surface energy versus $\delta q$ and $\delta \varphi$. For $\delta q$ larger than $\Delta$, the dependence on $\delta \varphi$ disappears, but for small $\delta q$ and $\delta \varphi$ the energy is near parabolic and given by the approximation above.
This expression, by assuming a rigid DW profile, overestimates the coupling energy and can be expected to slightly underestimate $\delta q$ and $\delta \varphi$.

This collective coordinates model covers a very large number of cases that are not always analytically solvable. We consider here the steady-state motion of the DW in translational or asymptotic precessional regimes. By assuming the bounded state of the two DWs, $\dot{\left<\delta\varphi\right>}\approx0$ and $\dot{\left<\delta q\right>}\approx0$. This allows us to obtain analytically expressions for $v$, $\varphi$, $\delta \varphi$ and $\delta q$ for all driving forces (field, STT, or SOT) in the two considered example systems, the CoGd film and the SAF, shown in Table~\ref{tab:2qPhi}. In these solutions, we apply the previously-discussed assumptions for CoGd ($t_1=t_2=t$, $P_2=\beta_2=\theta_{\rm SHE 2}=0$, $D_2=0$), and for the SAF  ($t=t_1+t_2$, $\theta_{\rm SHE 2}=0$, $D_2=0$, same values in layers `1' and `2' for the other parameters), and we have neglected the dipolar-induced planar anisotropy $K_{\rm BN}$.

Interestingly, we found that this model gives the same steady-state results ($v$, $\varphi$ and $\dot{\varphi}$) as the simpler $q$-$\varphi$ model with effective parameters for all driving forces (see Table~\ref{tab:2qPhi}). This shows the $q$-$\varphi$ model, that was obtained in the strong-coupling approximation of perfect antiparallel alignment, is valid even for finite coupling. Furthermore, it shows that the DW dynamics are independent of $J_{\rm AF}$.  \\

\begin{table*}[ht]
\centering
\begin{tabular*}{\textwidth}{l|@{\hskip 0.2cm}c@{\hskip 0.2cm}|@{\hskip 0.2cm}c}
    & \textbf{CoGd} & \textbf{SAF}\\
    \hline

	\textbf{Field-driven}, &
	$v=v_{\eff}$,  $\varphi=\varphi_{\eff}$ &
	$v=v_{\eff}$,  $\varphi=\varphi_{\eff}$ \\
	Translational & $\delta q= - \Delta \mu_0 H \frac{L_{\alpha 1} M_2 + L_{\alpha 2} M_1}{L_\alpha} / J_{\rm AF} $ &
	$\delta q= - \Delta \mu_0 H M_1 \frac{2 t_1 t_2}{t} /J_{\rm RKKY} $   \\
	& $\delta \varphi= -v L_2 /(\Delta J_{\rm AF}) $ &
	$\delta \varphi= -v L_1 t_2/(\Delta J_{\rm RKKY}) $   \\

	\hline
	\textbf{Field-driven}, &
	$v=v_{\eff}$, $\dot{\varphi}=\dot{\varphi}_{\eff}$ &
	$v=v_{\eff}$, $\dot{\varphi}=\dot{\varphi}_{\eff}$\\
	Asympt.  & $\delta q= -\Delta \mu_0 H \frac{L_S(L_1 M_2-L_2 M_1)+L_\alpha (L_{\alpha 1} M_2 +L_{\alpha 2} M_1 )}{L_S^2+L_\alpha^2}/J_{\rm AF} $ &
	$\delta q= -\Delta \mu_0 H M_1   \frac{L_\alpha^2}{L_S^2+L_\alpha^2} \frac{ 2 t_1 t_2}{t}/J_{\rm RKKY}$ \\
	Precessional & $\delta \varphi= -v  \frac{L_1 L_2(\alpha_1+\alpha_2)}{L_\alpha} / (\Delta J_{\rm AF}) $ &
	$\delta \varphi= -\mu_0 H M_1 \frac{L_S L_\alpha}{L_S^2+L_\alpha^2} \frac{2 t_1 t_2}{t}/J_{\rm RKKY}$ \\

	\hline
	\textbf{STT-driven}, &
	$v=v_{\eff}$, $\varphi=\varphi_{\eff}$ &
	$v=v_{\eff}$, $\varphi=\varphi_{\eff}$\\
     Translational & $\delta q = -v L_{\alpha 2}/J_{\rm AF} $ &
    $\delta q = 0 $    \\
	& $\delta \varphi = -v L_2 /(\Delta J_{\rm AF}) $  &
	$\delta \varphi = (v-u_1) L_{\alpha 1} t_2 /J_{\rm RKKY} $\\

	\hline
	\textbf{STT-driven}, &
	$v=v_{\eff}$, $\dot{\varphi}=\dot{\varphi}_{\eff}$ &
	$v=v_{\eff}$, $\dot{\varphi}=\dot{\varphi}_{\eff}$\\
	 Asympt.  & $\delta q = -v L_2 \frac{\beta L_2 \alpha_2^2-\beta L_{S} + L_1 (\beta \alpha_1 \alpha_2+\alpha_1+\alpha_2)}{\beta L_\alpha + L_{S}}/J_{\rm AF} $ &
	$\delta q = L_S u_1 (\beta_1-\alpha_1)\frac{L_1^2}{L_S^2+L_\alpha^2} \frac{2 t_1 t_2}{t} /J_{\rm RKKY} $\\
	 Precessional &  $\delta \varphi = -v L_2 \frac{L_{S}-\alpha_2 L_\alpha + \beta L_1 (\alpha_1+\alpha_2) }{\beta L_\alpha + L_{S}}/(\Delta J_{\rm AF}) $ &
	$\delta \varphi =L_{\alpha 1} u_1 (\alpha_1-\beta_1)\frac{L_1^2}{L_S^2+L_\alpha^2} \frac{2 t_1 t_2}{t} /(\Delta J_{\rm RKKY})$ \\

	\hline
	\textbf{SOT-driven} &
	$v=v_{\eff}$, $ \varphi=\varphi_{\eff}$  &
	$v=v_{\eff}$, $ \varphi=\varphi_{\eff}$  \\
	& $\delta q= -v L_{\alpha 2}/J_{\rm AF} $ &
	$\delta q= \pm v L_{\alpha 1} t_2/ J_{\rm RKKY} $ \\
	& $\delta \varphi= -v L_2 /(\Delta J_{\rm AF})$ &
	$\delta \varphi= \pm v L_1 t_2/(\Delta J_{\rm RKKY}) $\\
	\hline
\end{tabular*}
\caption{Collective coordinates model ($q$, $\varphi$, $\delta q$, $\delta\varphi$) solutions for a DW driven by field, STT or SOT in the studied CoGd (left column) and SAF cases (right column). In the CoGd, $D_2=0$, and current-induced torques only in the TM sublattice `1' ($\theta_{\rm SHE 2}=P_2=\beta_2=0$). In the SAF, the two layers have the same material parameters except $D_2=0$, and with current-induced torques only in the sublattice '1' ($\theta_{\rm SHE 2}=P_2=\beta_2=0$). In both cases, DMI is considered dominant relative to the dipolar-induced anisotropy ($K_{BN}=0$). $\delta q$ and $\delta\varphi$ are considered constant in time. $v_{\eff}$, $\varphi_{\eff}$ are the solutions of the $q$-$\varphi$ model with effective parameters (Table~\ref{tab:qphi_noDivergences}). The material parameters with no indices are the effective parameters defined in Section~\ref{sect:Models}.
}
\label{tab:2qPhi}
\end{table*}

\begin{figure}
    \centering
    \includegraphics[width=\columnwidth]{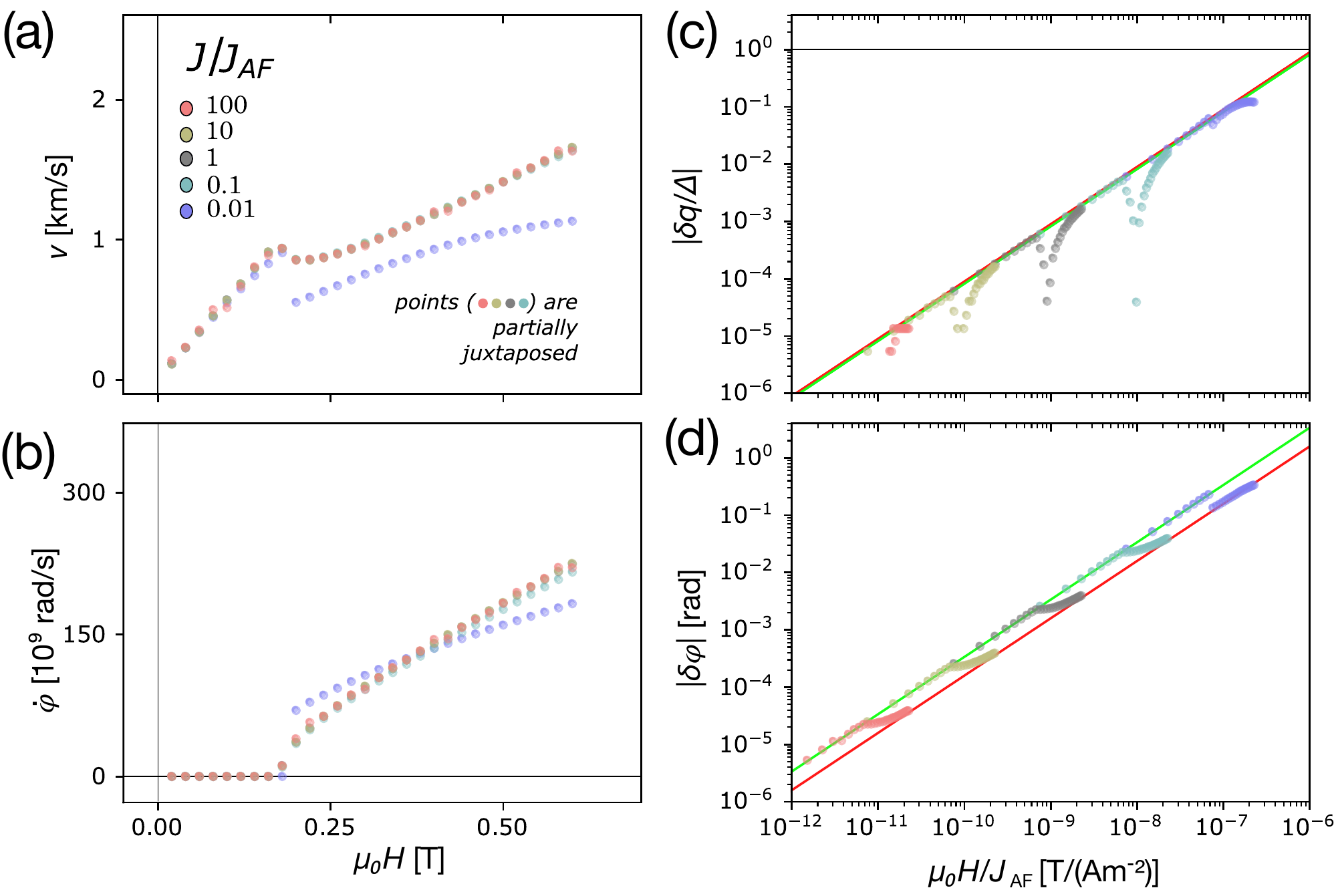}
	\caption[]{Effect of inter-sublattice angle on the velocity and precession rate of a field-driven DW in CoGd (\%Co=80\%, $m_S=0.07$, $l_S=0.018$).
	\textbf{(a, b)} Comparison between the (a) velocity and (b) precession rate versus the applied field obtained with different values of coupling (ranging from $J/J_{\rm AF}=100$ to 0.01). Many points are juxtaposed.
	\textbf{(c)} $\delta q/\Delta$ and \textbf{(d)} $\delta\varphi$ versus the ratio $H/J_{\rm AF}$ for the same simulations as in (a). The lines correspond to the linear relations predicted by the model (Table~\ref{tab:2qPhi}) for the translational (green) and asymptotic precessional (red) regimes.
	}
	\label{fig:CouplingDWdyn}
\end{figure}

\subsection{Micromagnetic simulations}
To verify this finding, we have repeated the micromagnetic simulations of a field-driven DW in CoGd with different values of the coupling constant $J_{\rm AF}$, from $100\times$ to $0.01\times$ the previously used value (in Table~\ref{tab:pars}).
For values of $J_{\rm AF}$ smaller than $0.01 \times J_{\rm AF}$, the DWs decoupled. Fig.~\ref{fig:CouplingDWdyn}a,b shows the velocity $v$ and precession rate $\dot{\varphi}$ versus $H$ obtained with the various values of $J_{\rm AF}$ for $\% Co=80\%$ ($m_S=0.07$). The colour of the points indicates the value of $J_{\rm AF}$.
For the largest couplings (100$\times$ to 0.1$\times$), $v$ and  $\dot{\varphi}$ coincide and are independent of $J_{\rm AF}$, as predicted by the model. Significant deviations occur only for the lowest couplings at the highest fields, where the DW structure is highly distorted.
Similar results were obtained for the other driving forces (not shown).
\\

\begin{figure}
    \centering
    \includegraphics[width=\columnwidth]{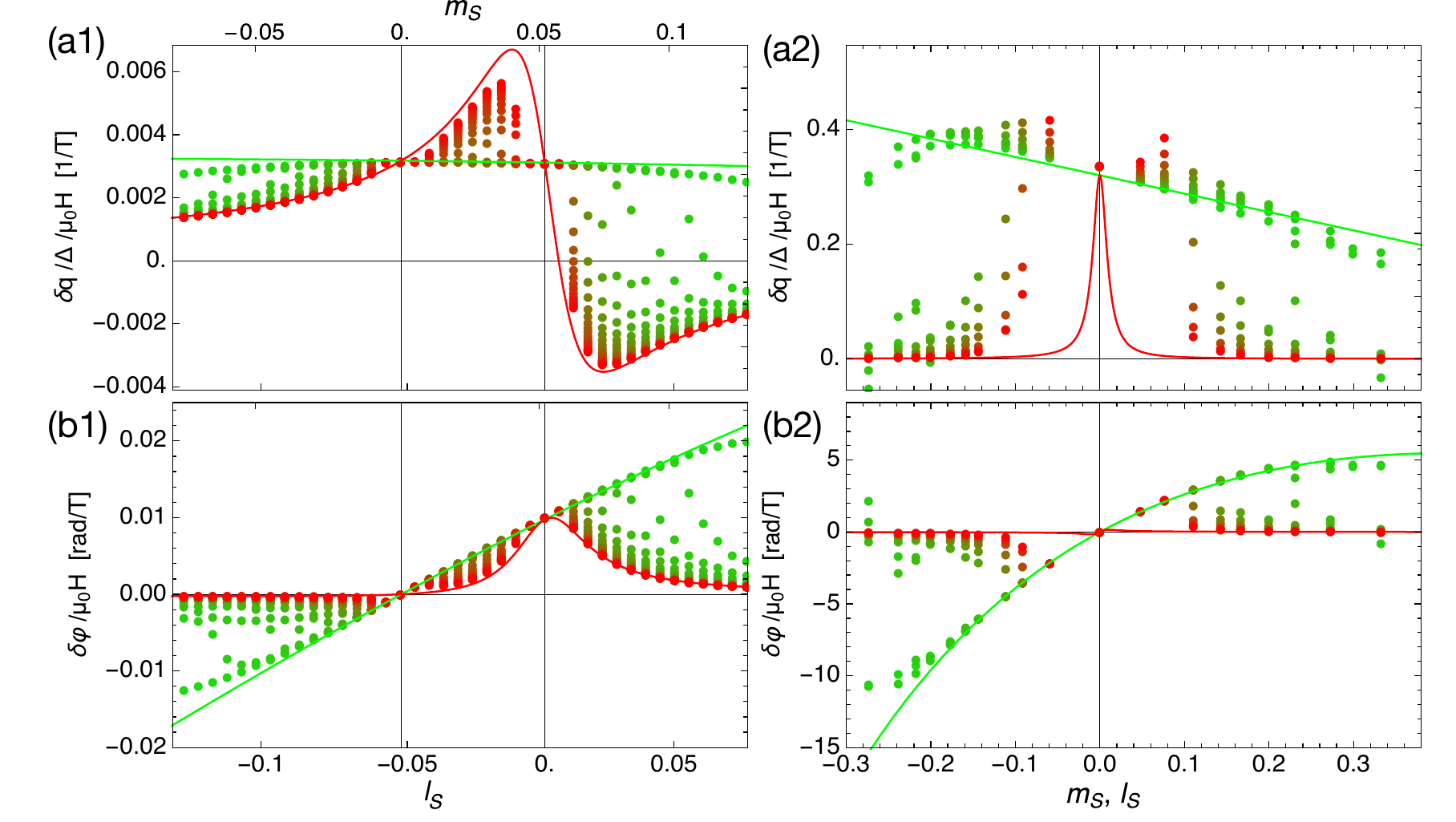}
	\caption[]{Variation of inter-lattice distortions,
	$\delta q$ and $\delta \varphi$, with $m_S$ of a field-driven DW in CoGd (a1,b1) and in the SAF (a2,b2).
	\textbf{(a1,2)} $\delta q/(H/J_{\rm AF})$ and
	\textbf{(b1,2)} $\delta\varphi/(H/J_{\rm AF})$ versus $l_S$. The lines are the analytical laws given by the collective coordinates model for the translational (green) and precessional (red) regimes (Table~\ref{tab:2qPhi}). The colour code of the points is the same as in Fig.\ref{fig:DWM_H}. }
	\label{fig:Couplingdqdphi}
\end{figure}

The model predicts that the internal distortions, $\delta q$ and $\delta\varphi$, are proportional to the ratio of the driving force and the inter-exchange coupling (e.g. $H/J_{\rm AF}$ for the field-driven DW), and so tend to zero for low driving forces or large coupling.
Figs.~\ref{fig:CouplingDWdyn}c and d show $\delta q/\Delta$ and $\delta\varphi$ versus $H/J_{\rm AF}$, for different $J_{\rm AF}$ (for $\% Co=80\%$).
Although $\delta q$ is sometimes neglected (e.g.~\cite{Martinez2019}), the magnitudes of $\delta \varphi$ and $\delta q/\Delta$ are comparable, and therefore produce comparable contributions to the system energy (Eq.~\ref{eq:Sigmaex}).

These distortions do not affect the DW velocity nor its precession, except at very low $J_{\rm AF}$. The velocity with $100\times J_{\rm AF}$ and with $0.1\times$ are the same, even if $\delta q$ and $\delta \varphi$ change by 4 orders-of-magnitude. Nevertheless, these distortions may be significant in the transient response~\cite{Chauleau2010}.
The linear relation of $\delta q$ and $\delta\varphi$ with $H/J_{\rm AF}$ predicted by the collective coordinates model can be clearly seen.
As expected, the proportionality factors ($\frac{\delta q}{\Delta} / \frac{\mu_0 H}{J_{\rm AF}}$ and $\frac{\delta \varphi}{\pi} / \frac{\mu_0 H}{J_{\rm AF}}$) are different for the translational and precessional regimes.
The proportionality factors also vary with $l_S$ (shown in Figs.~\ref{fig:Couplingdqdphi}a1,b1) as predicted by the model (Table~\ref{tab:2qPhi}). The same observation holds also for the SAF (shown in Fig.~\ref{fig:Couplingdqdphi}a2,b2).

In previous studies (e.g. \cite{Yang2015,Krishnia2017}) the DW velocity was described as function of the ``exchange coupling torque'' proportional to $J_{AF} \cos(\delta \varphi)$. However, the presented results show that the exchange coupling torque is not a convenient parameter, as the DW velocity does not directly depend on $\delta \varphi$ nor on the coupling torque. The coupling torque, being symmetric on the two sub-lattices, applies no net force on the DW, and may only influence the DW dynamics by distorting its structure.   \\

These results validate the use of $q$-$\varphi$ model with effective parameters to interpret DW dynamics, even at moderate coupling strength, and the use of the collective coordinates model to estimate the distortions of the DW pair ($\delta q$ and $\delta\varphi$). Gomonay et al.~\cite{Gomonay2014} showed that the antiferromagnets can be described by a single effective equation describing the Néel vector field, with the anti-alignment distortion as a subordinate expression. Here, as well, the dynamics of $q$ and $\varphi$ do not depend on $\delta q$ and $\delta \varphi$. The current result, although stricter in application domain (1D DW dynamics instead of the general LLG), suggests that such a description may be true in the wider class of ferrimagnetic systems.

\section{Conclusion}

Using analytical and numerical methods, we have studied the dynamics of magnetic DWs in systems with two antiferromagnetically-coupled sub-lattices, in particular in ferrimagnets and SAFs, under different driving forces: magnetic field, STT, and interfacial SOT.
By comparing it to micromagnetic simulations, we have shown that the the \textit{q}-$\varphi$ model using effective parameters provides an unified, analytical description of DW dynamics in multi-lattice systems. Moreover, we have shown that this description is valid even far from the strong-coupling limit, where the sublattices are not perfectly antiparallel. The same approach, using effective parameters, could be applied to other systems (e.g. multi-layered systems) or other magnetic textures (e.g. vortices or skyrmions).

In tunable systems, where the ratio of the sublattice moments can be changed, the models predict that the DW dynamics can differ significantly from the ferromagnetic case in the vicinity of the compensation points (ACP and MCP). Depending on the driving force, the DW may show features with great interest for applications, such as a enhanced mobility or suppression of precession. As we have discussed, some of these features can be seen in recently reported results on DW dynamics in systems like ferrimagnets or synthetic anti-ferromagnets, while other predicted features are yet to be observed.

\begin{acknowledgements}
We acknowledge fruitful discussions with Stanislas Rohart and André Thiaville.

S. K., and E. H. acknowledge public grant overseen by the ANR as part of the \textit{``Investissements d’Avenir''} programme (Labex NanoSaclay, reference: ANR-10-LABX-0035) for the FEMINIST project. S.K. acknowledges a public grant from PIAF ANR-17-CE09-0030. J.S. and A.M. acknowledge funding from Labex Nanosaclay Flagship project SPICY.
\end{acknowledgements}

\FloatBarrier
\appendix

\section{The $q$-$\varphi$ model \label{sect:Appendix_qPhi} }

The $q$-$\varphi$ model is deduced from the LLG equation by assuming a Bloch DW profile, where the magnetisation in spherical coordinates, $(\Theta(x),\phi(x))$, follows
\begin{align*}
    \phi(x)     &= \varphi \\
    \Theta(x)   &= 2 \arctan(\exp(x-q)/\Delta)
\end{align*}
where $\varphi$ is the in-plane angle of the magnetisation of the DW (independent of $x$), $q$ the position of the DW centre, and $\Delta$ the Bloch's DW width parameter (see Fig.~\ref{fig:2DWferri}c). The DW dynamics under driving forces (field $\vec{H}$ or electrical current $J$) can be described via the time-evolution of the two coupled collective coordinates ($q$ and $\varphi$):
    \begin{equation}
    \begin{cases}
    \alpha \frac{\dot{q}}{\Delta}+\dot{\varphi} & = \gamma_0 H_z   -\frac{\partial_q \sigma_{\rm ex}}{2 L_S}  +\frac{\pi}{2} \frac{\tau_{DL}}{L_S} \cos\varphi +\beta \frac{u}{\Delta}\\
    \frac{\dot{q}}{\Delta}-\alpha\dot{\varphi} & = \gamma_0 H_{\rm BN} \frac{\sin(2\varphi)}{2}  +\frac{\pi}{2} \gamma_0   (H_{\rm DMI}+H_x) \sin\varphi -\\
     & -(\gamma_0 H_y +\frac{\tau_{FL}}{L_S}) \cos\varphi   +\frac{\partial_{\varphi}\sigma_{\rm ex}}{2\Delta L_S}  + \frac{u}{\Delta}\\
    \end{cases}
    \label{eq:qphi}
    \end{equation}
The term $\sigma_{\rm ex}$ accounts for the other energy terms of the system  (such as the inter-lattice exchange coupling in sect.~\ref{sect:Low_Coupling}), $H_{\rm DMI} = \frac{D/\Delta}{\mu_0 M_S}$ accounts for the DMI, and $H_{\rm BN} = \frac{2 K_{\rm BN}}{\mu_0 M_S}$ accounts for the in-plane (Bloch/Néel) dipolar-induced anisotropy. For thin films with spatially-merged sublattices, $K_{\rm BN} \approx \frac{\ln 2}{2 \pi} \frac{t}{\Delta} \mu_0 M_S^2$ \cite{Tarasenko1998}. For systems with spatially-separated sublattices the same approximation would yield $K_{\rm BN} \approx \frac{\ln 2}{2 \pi}  \mu_0 (t_1^2 M_1^2+ t_2^2 M_2^2)/(t \Delta)$, but the dipolar interactions between the DWs in the two layers will decrease it by orders of magnitude~\cite{Hrabec2018}  and so $K_{\rm BN}$ may be neglected in the presence of DMI.
In agreement with this description, two steady-state DW propagation regimes can exist~\cite{Schryer1974}: the translational regime at low drive, with a constant velocity and fixed DW angle ($\dot{\varphi}=0$), and the precessional regime at high drive, with oscillating velocity and oscillating DW angle. The two regimes are
separated by the Walker threshold~\cite{Schryer1974, Malozemoff1979, Mougin2007, Thiaville2012}, which is the maximum driving force that still satisfies the condition $\dot{\varphi}=0$. In the limit of high drive, the DW approaches the asymptotic precessional regime, where $\dot{\varphi} \gg 0$ and it is possible to consider that  $\left\langle \sin \varphi \right\rangle \approx \left\langle \cos \varphi \right\rangle \approx 0$. Note that for a DW driven by SHE alone, the precessional regime does not exist~\cite{Thiaville2012}.  The velocity $v=\dot{q}$, the internal angle $\varphi$ and the precession rate $\dot{\varphi}$ for a DW driven by field, SOT or STT in the translational and the asymptotic precessional regimes, deduced from eq.~\ref{eq:qphi}, are shown in Table~\ref{tab:Classical_qphi}. This model can also be applied to DWs in in-plane magnetised systems, with the due adaptations of the effective fields.

\begin{table*}[ht]
\centering
\begin{tabular*}{\textwidth}{l@{\extracolsep{\fill}}l|c|ll}
	\multicolumn{2}{l}{\textbf{Field-driven}} \\
	\multicolumn{2}{c}{Translational regime}  &
	\multicolumn{1}{c}{Walker threshold} &	\multicolumn{2}{c}{Asympt. Precessional regime} \\	\hline
	$ v = \frac{\Delta}{\alpha} \gamma_0 H  $ &
	$\varphi = \begin{cases} \frac{1}{2} \arcsin\left( H/H_{W}
	\right)\\  \arcsin\left(
	H/H_{W}
	\right) \end{cases} $ &
	$\left| H_W \right| = \begin{cases}  \frac{1}{2} \left| \alpha H_{\rm BN} \right| \\  \frac{\pi}{2} \left| \alpha H_{\rm DMI} \right|  \end{cases} $     &
	$v = \frac{\Delta \alpha }{1+\alpha^2}\gamma_0 H$ &
	$\dot{\varphi}  =     \frac{1}{1+\alpha^2} \gamma_0 H$ \\

	\hline
	\multicolumn{5}{l}{ } \\
	\multicolumn{5}{l}{\textbf{STT-driven}} \\
	\multicolumn{2}{c}{Translational regime}  &
    \multicolumn{1}{c}{Walker threshold} &	\multicolumn{2}{c}{Asympt. Precessional regime} \\	\hline
	$v= \frac{\beta }{\alpha} u$ &
	$\varphi = \begin{cases} \frac{1}{2} \arcsin\left(
	u/u_W
	\right)\\   \arcsin\left(
	u/u_W
	\right) \end{cases} $ &
	$\left| u_W \right| = \begin{cases}  \frac{1}{2}  \left| \frac{ \alpha \Delta \gamma_0 H_{\rm BN} }{\beta-\alpha}\right|\\ \frac{\pi}{2}  \left| \frac{\alpha \Delta \gamma_0 H_{\rm DMI} }{\beta-\alpha}\right| \end{cases} $       &
	$v =  \frac{1+\alpha\beta}{1+\alpha^2}  u$ &
	$\dot{\varphi}=\frac{\beta-\alpha}{1+\alpha^2} \frac{u}{\Delta}$ \\

	\hline
	\\
	\multicolumn{2}{l}{\textbf{SOT-driven}} \\ \hline
	$ v = \frac{\Delta}{\alpha} \frac{\pi}{2} \frac{\tau_{\rm SHE}}{L_S} \cos\varphi$ &
	$\varphi = \arctan\left(\frac{\tau_{\rm SHE}}{\alpha M_S \mu_0 H_{\rm DMI}}\right)$ &  \multicolumn{1}{c}{--}  & \multicolumn{2}{c}{--} \\
	\hline
\end{tabular*}
\caption[]{$q-\varphi$ model solutions (eq.~\ref{eq:qphi}) for $v$, $\varphi$, $\dot{\varphi}$ and the Walker threshold of a DW in a ferromagnet with perpendicular anisotropy driven by field, STT or SOT in translational and asymptotic precessional regimes~\cite{Schryer1974,Mougin2007,Thiaville2012}. The top and bottom expressions for $\varphi$ and Walker thresholds refer to the case of dominant dipolar-induced Bloch-Néel anisotropy ($K_{\rm BN}$) or dominant DMI ($D$). Here, $H_x=H_y=H_{\rm FL}=0$. $H_{\rm BN}=\frac{2K_{\rm BN}}{\mu_0 M_S}$, $H_{\rm DMI}= \frac{ D/\Delta }{\mu_0 M_S}$. }
\label{tab:Classical_qphi}
\end{table*}

\section{Micromagnetic energy terms \label{sect:umagEnerg}}

The usual energy volumic density terms due to Zeeman energy, uniaxial anisotropy, exchange stiffness, DMI, and demagnetising field ($\vec{H}_{\rm demag}$) are
\begin{align*}
    U_{\rm Zeeman}  &= -\mu_0 M_S \vec{m} \cdot \vec{H}_0   \\
                U_K &= -K(\vec{m}\cdot\vec{z})^2             \\
        U_{\rm ex.} &= A |\nabla \vec{m}|^2                 \\
            U_{DMI} &= D[ m_z \vec{\nabla}\cdot \vec{m} - (\vec{m} \cdot \vec{\nabla}) m_z]                              \\
    U_{\rm demag.}  & =-\frac{1}{2}\mu_0 M_S \vec{m} \cdot \vec{H}_{\rm demag}
\end{align*}

In double lattice systems, the inter-sublattice exchange coupling can be expressed as $  U_{\rm 12}=-J_{\rm AF} \vec{m}_1 \cdot \vec{m}_2 $. In the SAF, the RKKY coupling is an interface effect. As we only consider the limit of thin films in this article (i.e., $\partial_z\vec{m}\approx 0$), the same form of $U_{\rm 12}$ can be used, with $J_{\rm RKKY}/t$ replacing $J_{\rm AF}$, $J_{\rm RKKY}$ being an areal energy density.

\begin{table*}[ht]
\begin{tabular}{p{3cm}p{2cm}p{2cm}p{10cm}}
\textbf{Symbol}         & \textbf{Units}    & \textbf{Usage}   & \textbf{Description}                                \\ \hline
$\vec{m}$, $\vec{M}$    & --, A/m           & Eqs. \ref{eq_llg_old}, \ref{eq:LLG_Full} & Normalised magnetisation and magnetisation, $\vec{M}=M_S\vec{m}$. \\
$M_S$, $M_1$, $M_2$     & A/m               & Eq. \ref{eq:Eff_Params} & Spontaneous magnetisation. \\
$L_S$, $L_1$, $L_2$     & Kg/(ms)           & Eqs. \ref{eq:LLG_Full},\ref{eq:Eff_Params} & Angular momentum density. \\
$m_S$                   & --                & Eq. \ref{eq:msls} & normalised magnetisation. \\
$l_S$                   & --                & Eq. \ref{eq:msls} & normalised angular momentum. \\
$\gamma$, $\gamma_0$    & s$^{-1}/$T, s$^{-1}/$(A/m) & Eq. \ref{eq_llg_old} & Gyromagnetic ratio. $\gamma_0= \mu_0\gamma =\mu_0 M_S/L_S$.     \\
$g$                     & --                & & Landé g-factor, $\gamma=g\mu_B/\hbar$, where $\mu_B$ is the Bohr magneton. \\

$\alpha$                & --                & Eqs. \ref{eq_llg_old}, \ref{eq:Eff_Params}& Gilbert damping parameter, $\alpha=L_\alpha/L_S$. \\
$L_\alpha$, $L_{\alpha 1}$,
$L_{\alpha 2}$          & sJ/m$^{3}$        & Eqs. \ref{eq:LLG_Full},\ref{eq:Eff_Params} & Product $\alpha L_S$.       \\
$t$, $t_1$, $t_2$ & m   & & Film (sublattice) thickness.\\
$\oldvec{\mathbfcal{H}}$ & A/m              & & Effective field, $\oldvec{\mathbfcal{H}}\equiv\frac{-1}{\mu_0 M_S} \dmU$. \\
$\dmU$                  & J/m$^{3}$         &eq. \ref{eq:LLG_Full} & Variational derivative of  $U(\vec{m}(\vec{r}))$.    \\
$U$, $U_1$, $U_2$       & J/m$^{3}$         & Sect. \ref{sect:umagEnerg} & Total energy (density). $U=U_{\rm Zeeman}+U_K + U_{\rm ex} + U_{\rm DMI} + U_{\rm demag}$ (due to Zeeman, anisotropy, exchange, Dzyaloshinskii-Moriya, and dipolar effects.)     \\
$\sigma_{\rm ex}$   & J/m$^{2}$     & Sect. \ref{sect:Appendix_qPhi} & Interlattice coupling energy per DW surface. \\
$K$, $K_{\rm 1}$, $K_{\rm 2}$,
$K_{\rm dipolar}$       & J/m$^{3}$         & Sect.~\ref{sect:umagEnerg} & Effective, internal and dipolar-induced uniaxial anisotropy. \\
$K_{\rm BN}$            & J/m$^{3}$         & Sect. \ref{sect:Appendix_qPhi} & Dipolar-induced in-plane (Bloch/Néel) DW anisotropy. \\
$A$, $A_1$, $A_2$       & J/m               & Sect.~\ref{sect:umagEnerg} & Exchange stiffness.\\
$D$, $D_1$, $D_2$         & J/m$^{2}$         & Sect.~\ref{sect:umagEnerg} & DMI parameter.\\
$J_{\rm AF}$            & J/m$^{3}$         & Sect.~\ref{sect:umagEnerg} & Coupling between sublattices (volumic energy density).\\
$J_{\rm RKKY}$          & J/m$^{2}$         & Sect.~\ref{sect:umagEnerg} & Coupling between the two layers of the SAF (areal energy density). $J_{\rm AF}=J_{\rm RKKY}/t$. \\
$\vec{\tau}$, $\vec{\tau}_{\rm STT}$,
$\vec{\tau}_{\rm SOT}$, $\vec{\tau}_{\rm FL}$,
$\vec{\tau}_{DL}$       & Nm/m$^3$          & Eq.\ref{eq:LLG_Full}  & Non-conservative torques, which include the STT, SOT, field-like SOT, and damping-like SOT.  \\
$\tau_{\rm SHE}$        & Nm/m$^3$          & & SHE-induced SOT. $\tau_{\rm SHE}=J\theta_{\rm SHE} \hbar/(2et)$. \\
$\vec{u}$               & m/s               & & STT parameter. $L_S \vec{u} = PJ\hbar/(2e)\vec{e}_J$.\\
$\beta$, $\beta_1$, $\beta_2$  & --         & Eq. \ref{eq:Eff_Params_J} & STT non-adiabatic parameter.\\
$P$, $P_1$, $P_2$       & --                & Eq. \ref{eq:Eff_Params_J} & Current spin polarisation (STT).\\
$\theta_{\rm SHE}$, $\theta_{\rm SHE1}$,
$\theta_{\rm SHE2}$     &--                 & Eq. \ref{eq:Eff_Params_J} & SHE angle.\\
$v$                     & m/s               & & DW velocity. \\
$q$                     & m                 & Sect.~\ref{sect:Appendix_qPhi}   & DW position (\textit{q}-$\varphi$ model.) \\
$\varphi$               & (rad)             & Sect.~\ref{sect:Appendix_qPhi}   & In-plane angle of DW magnetisation (\textit{q}-$\varphi$ model.)\\
$\Delta$                & m                 & Sect.~\ref{sect:Appendix_qPhi}   &  Bloch DW width parameter.\\
$\delta q$              & m                 & Sect. \ref{sect:Model_Collective_Finite} & DW position difference. \\
$\delta \varphi$        & (rad)             & Sect. \ref{sect:Model_Collective_Finite} & DW in-plane angle difference.\\
$H_W$                   & A/m               & & Walker threshold (field-driven.) \\
$J_W$                   & A/m$^{2}$         & & Walker threshold (STT-driven.) \\
\end{tabular}
\caption{List of symbols used in the text. Parameters with indices `1' and `2' refer to the sublattice parameters.}
\label{tab:glossary}
\end{table*}

\FloatBarrier

\bibliography{BibtexManually}

\begin{thebibliography}{48}%
\makeatletter
\providecommand \@ifxundefined [1]{%
 \@ifx{#1\undefined}
}%
\providecommand \@ifnum [1]{%
 \ifnum #1\expandafter \@firstoftwo
 \else \expandafter \@secondoftwo
 \fi
}%
\providecommand \@ifx [1]{%
 \ifx #1\expandafter \@firstoftwo
 \else \expandafter \@secondoftwo
 \fi
}%
\providecommand \natexlab [1]{#1}%
\providecommand \enquote  [1]{``#1''}%
\providecommand \bibnamefont  [1]{#1}%
\providecommand \bibfnamefont [1]{#1}%
\providecommand \citenamefont [1]{#1}%
\providecommand \href@noop [0]{\@secondoftwo}%
\providecommand \href [0]{\begingroup \@sanitize@url \@href}%
\providecommand \@href[1]{\@@startlink{#1}\@@href}%
\providecommand \@@href[1]{\endgroup#1\@@endlink}%
\providecommand \@sanitize@url [0]{\catcode `\\12\catcode `\$12\catcode
  `\&12\catcode `\#12\catcode `\^12\catcode `\_12\catcode `\%12\relax}%
\providecommand \@@startlink[1]{}%
\providecommand \@@endlink[0]{}%
\providecommand \url  [0]{\begingroup\@sanitize@url \@url }%
\providecommand \@url [1]{\endgroup\@href {#1}{\urlprefix }}%
\providecommand \urlprefix  [0]{URL }%
\providecommand \Eprint [0]{\href }%
\providecommand \doibase [0]{https://doi.org/}%
\providecommand \selectlanguage [0]{\@gobble}%
\providecommand \bibinfo  [0]{\@secondoftwo}%
\providecommand \bibfield  [0]{\@secondoftwo}%
\providecommand \translation [1]{[#1]}%
\providecommand \BibitemOpen [0]{}%
\providecommand \bibitemStop [0]{}%
\providecommand \bibitemNoStop [0]{.\EOS\space}%
\providecommand \EOS [0]{\spacefactor3000\relax}%
\providecommand \BibitemShut  [1]{\csname bibitem#1\endcsname}%
\let\auto@bib@innerbib\@empty
\bibitem [{\citenamefont {Manchon}\ \emph {et~al.}(2019)\citenamefont
  {Manchon}, \citenamefont {{\v{Z}}elezn{\'{y}}}, \citenamefont {Miron},
  \citenamefont {Jungwirth}, \citenamefont {Sinova}, \citenamefont {Thiaville},
  \citenamefont {Garello},\ and\ \citenamefont {Gambardella}}]{Manchon2019}%
  \BibitemOpen
  \bibfield  {author} {\bibinfo {author} {\bibfnamefont {A.}~\bibnamefont
  {Manchon}}, \bibinfo {author} {\bibfnamefont {J.}~\bibnamefont
  {{\v{Z}}elezn{\'{y}}}}, \bibinfo {author} {\bibfnamefont {I.~M.}\
  \bibnamefont {Miron}}, \bibinfo {author} {\bibfnamefont {T.}~\bibnamefont
  {Jungwirth}}, \bibinfo {author} {\bibfnamefont {J.}~\bibnamefont {Sinova}},
  \bibinfo {author} {\bibfnamefont {A.}~\bibnamefont {Thiaville}}, \bibinfo
  {author} {\bibfnamefont {K.}~\bibnamefont {Garello}},\ and\ \bibinfo {author}
  {\bibfnamefont {P.}~\bibnamefont {Gambardella}},\ }\bibfield  {title}
  {\bibinfo {title} {{Current-induced spin-orbit torques in ferromagnetic and
  antiferromagnetic systems}},\ }\href
  {https://doi.org/10.1103/RevModPhys.91.035004} {\bibfield  {journal}
  {\bibinfo  {journal} {Reviews of Modern Physics}\ }\textbf {\bibinfo {volume}
  {91}},\ \bibinfo {pages} {035004} (\bibinfo {year} {2019})}\BibitemShut
  {NoStop}%
\bibitem [{\citenamefont {Avci}\ \emph {et~al.}(2019)\citenamefont {Avci},
  \citenamefont {Rosenberg}, \citenamefont {Caretta}, \citenamefont
  {B{\"{u}}ttner}, \citenamefont {Mann}, \citenamefont {Marcus}, \citenamefont
  {Bono}, \citenamefont {Ross},\ and\ \citenamefont {Beach}}]{Avci2019}%
  \BibitemOpen
  \bibfield  {author} {\bibinfo {author} {\bibfnamefont {C.~O.}\ \bibnamefont
  {Avci}}, \bibinfo {author} {\bibfnamefont {E.}~\bibnamefont {Rosenberg}},
  \bibinfo {author} {\bibfnamefont {L.}~\bibnamefont {Caretta}}, \bibinfo
  {author} {\bibfnamefont {F.}~\bibnamefont {B{\"{u}}ttner}}, \bibinfo {author}
  {\bibfnamefont {M.}~\bibnamefont {Mann}}, \bibinfo {author} {\bibfnamefont
  {C.}~\bibnamefont {Marcus}}, \bibinfo {author} {\bibfnamefont
  {D.}~\bibnamefont {Bono}}, \bibinfo {author} {\bibfnamefont {C.~A.}\
  \bibnamefont {Ross}},\ and\ \bibinfo {author} {\bibfnamefont {G.~S.~D.}\
  \bibnamefont {Beach}},\ }\bibfield  {title} {\bibinfo {title}
  {{Interface-driven chiral magnetism and current-driven domain walls in
  insulating magnetic garnets}},\ }\href
  {https://doi.org/10.1038/s41565-019-0421-2} {\bibfield  {journal} {\bibinfo
  {journal} {Nature Nanotechnology}\ }\textbf {\bibinfo {volume} {14}},\
  \bibinfo {pages} {561} (\bibinfo {year} {2019})}\BibitemShut {NoStop}%
\bibitem [{\citenamefont {Jungwirth}\ \emph {et~al.}(2016)\citenamefont
  {Jungwirth}, \citenamefont {Marti}, \citenamefont {Wadley},\ and\
  \citenamefont {Wunderlich}}]{Jungwirth2016}%
  \BibitemOpen
  \bibfield  {author} {\bibinfo {author} {\bibfnamefont {T.}~\bibnamefont
  {Jungwirth}}, \bibinfo {author} {\bibfnamefont {X.}~\bibnamefont {Marti}},
  \bibinfo {author} {\bibfnamefont {P.}~\bibnamefont {Wadley}},\ and\ \bibinfo
  {author} {\bibfnamefont {J.}~\bibnamefont {Wunderlich}},\ }\bibfield  {title}
  {\bibinfo {title} {{Antiferromagnetic spintronics}},\ }\href
  {https://doi.org/10.1038/nnano.2016.18} {\bibfield  {journal} {\bibinfo
  {journal} {Nature Nanotechnology}\ }\textbf {\bibinfo {volume} {11}},\
  \bibinfo {pages} {231} (\bibinfo {year} {2016})}\BibitemShut {NoStop}%
\bibitem [{\citenamefont {Yang}\ \emph {et~al.}(2015)\citenamefont {Yang},
  \citenamefont {Ryu},\ and\ \citenamefont {Parkin}}]{Yang2015}%
  \BibitemOpen
  \bibfield  {author} {\bibinfo {author} {\bibfnamefont {S.-H.}\ \bibnamefont
  {Yang}}, \bibinfo {author} {\bibfnamefont {K.-S.}\ \bibnamefont {Ryu}},\ and\
  \bibinfo {author} {\bibfnamefont {S.}~\bibnamefont {Parkin}},\ }\bibfield
  {title} {\bibinfo {title} {{Domain-wall velocities of up to 750 m s−1
  driven by exchange-coupling torque in synthetic antiferromagnets}},\ }\href
  {https://doi.org/10.1038/nnano.2014.324} {\bibfield  {journal} {\bibinfo
  {journal} {Nature Nanotechnology}\ }\textbf {\bibinfo {volume} {10}},\
  \bibinfo {pages} {221} (\bibinfo {year} {2015})}\BibitemShut {NoStop}%
\bibitem [{\citenamefont {Hrabec}\ \emph {et~al.}(2018)\citenamefont {Hrabec},
  \citenamefont {Kři{\v{z}}{\'{a}}kov{\'{a}}}, \citenamefont {Pizzini},
  \citenamefont {Sampaio}, \citenamefont {Thiaville}, \citenamefont {Rohart},\
  and\ \citenamefont {Vogel}}]{Hrabec2018}%
  \BibitemOpen
  \bibfield  {author} {\bibinfo {author} {\bibfnamefont {A.}~\bibnamefont
  {Hrabec}}, \bibinfo {author} {\bibfnamefont {V.}~\bibnamefont
  {Kři{\v{z}}{\'{a}}kov{\'{a}}}}, \bibinfo {author} {\bibfnamefont
  {S.}~\bibnamefont {Pizzini}}, \bibinfo {author} {\bibfnamefont
  {J.}~\bibnamefont {Sampaio}}, \bibinfo {author} {\bibfnamefont
  {A.}~\bibnamefont {Thiaville}}, \bibinfo {author} {\bibfnamefont
  {S.}~\bibnamefont {Rohart}},\ and\ \bibinfo {author} {\bibfnamefont
  {J.}~\bibnamefont {Vogel}},\ }\bibfield  {title} {\bibinfo {title} {{Velocity
  Enhancement by Synchronization of Magnetic Domain Walls}},\ }\href
  {https://doi.org/10.1103/PhysRevLett.120.227204} {\bibfield  {journal}
  {\bibinfo  {journal} {Physical Review Letters}\ }\textbf {\bibinfo {volume}
  {120}},\ \bibinfo {pages} {227204} (\bibinfo {year} {2018})}\BibitemShut
  {NoStop}%
\bibitem [{\citenamefont {Hrabec}\ \emph {et~al.}(2017)\citenamefont {Hrabec},
  \citenamefont {Sampaio}, \citenamefont {Belmeguenai}, \citenamefont {Gross},
  \citenamefont {Weil}, \citenamefont {Ch{\'{e}}rif}, \citenamefont
  {Stashkevich}, \citenamefont {Jacques}, \citenamefont {Thiaville},\ and\
  \citenamefont {Rohart}}]{Hrabec2017}%
  \BibitemOpen
  \bibfield  {author} {\bibinfo {author} {\bibfnamefont {A.}~\bibnamefont
  {Hrabec}}, \bibinfo {author} {\bibfnamefont {J.}~\bibnamefont {Sampaio}},
  \bibinfo {author} {\bibfnamefont {M.}~\bibnamefont {Belmeguenai}}, \bibinfo
  {author} {\bibfnamefont {I.}~\bibnamefont {Gross}}, \bibinfo {author}
  {\bibfnamefont {R.}~\bibnamefont {Weil}}, \bibinfo {author} {\bibfnamefont
  {S.~M.}\ \bibnamefont {Ch{\'{e}}rif}}, \bibinfo {author} {\bibfnamefont
  {A.}~\bibnamefont {Stashkevich}}, \bibinfo {author} {\bibfnamefont
  {V.}~\bibnamefont {Jacques}}, \bibinfo {author} {\bibfnamefont
  {A.}~\bibnamefont {Thiaville}},\ and\ \bibinfo {author} {\bibfnamefont
  {S.}~\bibnamefont {Rohart}},\ }\bibfield  {title} {\bibinfo {title}
  {{Current-induced skyrmion generation and dynamics in symmetric bilayers}},\
  }\href {https://doi.org/10.1038/ncomms15765} {\bibfield  {journal} {\bibinfo
  {journal} {Nature Communications}\ }\textbf {\bibinfo {volume} {8}},\
  \bibinfo {pages} {15765} (\bibinfo {year} {2017})}\BibitemShut {NoStop}%
\bibitem [{\citenamefont {Kim}\ \emph {et~al.}(2017)\citenamefont {Kim},
  \citenamefont {Kim}, \citenamefont {Hirata}, \citenamefont {Oh},
  \citenamefont {Tono}, \citenamefont {Kim}, \citenamefont {Okuno},
  \citenamefont {Ham}, \citenamefont {Kim}, \citenamefont {Go}, \citenamefont
  {Tserkovnyak}, \citenamefont {Tsukamoto}, \citenamefont {Moriyama},
  \citenamefont {Lee},\ and\ \citenamefont {Ono}}]{Kim2017b}%
  \BibitemOpen
  \bibfield  {author} {\bibinfo {author} {\bibfnamefont {K.-J.}\ \bibnamefont
  {Kim}}, \bibinfo {author} {\bibfnamefont {S.~K.}\ \bibnamefont {Kim}},
  \bibinfo {author} {\bibfnamefont {Y.}~\bibnamefont {Hirata}}, \bibinfo
  {author} {\bibfnamefont {S.-H.}\ \bibnamefont {Oh}}, \bibinfo {author}
  {\bibfnamefont {T.}~\bibnamefont {Tono}}, \bibinfo {author} {\bibfnamefont
  {D.-H.}\ \bibnamefont {Kim}}, \bibinfo {author} {\bibfnamefont
  {T.}~\bibnamefont {Okuno}}, \bibinfo {author} {\bibfnamefont {W.~S.}\
  \bibnamefont {Ham}}, \bibinfo {author} {\bibfnamefont {S.}~\bibnamefont
  {Kim}}, \bibinfo {author} {\bibfnamefont {G.}~\bibnamefont {Go}}, \bibinfo
  {author} {\bibfnamefont {Y.}~\bibnamefont {Tserkovnyak}}, \bibinfo {author}
  {\bibfnamefont {A.}~\bibnamefont {Tsukamoto}}, \bibinfo {author}
  {\bibfnamefont {T.}~\bibnamefont {Moriyama}}, \bibinfo {author}
  {\bibfnamefont {K.-J.}\ \bibnamefont {Lee}},\ and\ \bibinfo {author}
  {\bibfnamefont {T.}~\bibnamefont {Ono}},\ }\bibfield  {title} {\bibinfo
  {title} {{Fast domain wall motion in the vicinity of the angular momentum
  compensation temperature of ferrimagnets}},\ }\href
  {https://doi.org/10.1038/nmat4990} {\bibfield  {journal} {\bibinfo  {journal}
  {Nature Materials}\ }\textbf {\bibinfo {volume} {16}},\ \bibinfo {pages}
  {1187} (\bibinfo {year} {2017})}\BibitemShut {NoStop}%
\bibitem [{\citenamefont {Siddiqui}\ \emph {et~al.}(2018)\citenamefont
  {Siddiqui}, \citenamefont {Han}, \citenamefont {Finley}, \citenamefont
  {Ross},\ and\ \citenamefont {Liu}}]{Siddiqui2018}%
  \BibitemOpen
  \bibfield  {author} {\bibinfo {author} {\bibfnamefont {S.~A.}\ \bibnamefont
  {Siddiqui}}, \bibinfo {author} {\bibfnamefont {J.}~\bibnamefont {Han}},
  \bibinfo {author} {\bibfnamefont {J.~T.}\ \bibnamefont {Finley}}, \bibinfo
  {author} {\bibfnamefont {C.~A.}\ \bibnamefont {Ross}},\ and\ \bibinfo
  {author} {\bibfnamefont {L.}~\bibnamefont {Liu}},\ }\bibfield  {title}
  {\bibinfo {title} {{Current-Induced Domain Wall Motion in a Compensated
  Ferrimagnet}},\ }\href {https://doi.org/10.1103/PhysRevLett.121.057701}
  {\bibfield  {journal} {\bibinfo  {journal} {Physical Review Letters}\
  }\textbf {\bibinfo {volume} {121}},\ \bibinfo {pages} {057701} (\bibinfo
  {year} {2018})}\BibitemShut {NoStop}%
\bibitem [{\citenamefont {Caretta}\ \emph {et~al.}(2018)\citenamefont
  {Caretta}, \citenamefont {Mann}, \citenamefont {B{\"{u}}ttner}, \citenamefont
  {Ueda}, \citenamefont {Pfau}, \citenamefont {G{\"{u}}nther}, \citenamefont
  {Hessing}, \citenamefont {Churikova}, \citenamefont {Klose}, \citenamefont
  {Schneider}, \citenamefont {Engel}, \citenamefont {Marcus}, \citenamefont
  {Bono}, \citenamefont {Bagschik}, \citenamefont {Eisebitt},\ and\
  \citenamefont {Beach}}]{Caretta2018}%
  \BibitemOpen
  \bibfield  {author} {\bibinfo {author} {\bibfnamefont {L.}~\bibnamefont
  {Caretta}}, \bibinfo {author} {\bibfnamefont {M.}~\bibnamefont {Mann}},
  \bibinfo {author} {\bibfnamefont {F.}~\bibnamefont {B{\"{u}}ttner}}, \bibinfo
  {author} {\bibfnamefont {K.}~\bibnamefont {Ueda}}, \bibinfo {author}
  {\bibfnamefont {B.}~\bibnamefont {Pfau}}, \bibinfo {author} {\bibfnamefont
  {C.~M.}\ \bibnamefont {G{\"{u}}nther}}, \bibinfo {author} {\bibfnamefont
  {P.}~\bibnamefont {Hessing}}, \bibinfo {author} {\bibfnamefont
  {A.}~\bibnamefont {Churikova}}, \bibinfo {author} {\bibfnamefont
  {C.}~\bibnamefont {Klose}}, \bibinfo {author} {\bibfnamefont
  {M.}~\bibnamefont {Schneider}}, \bibinfo {author} {\bibfnamefont
  {D.}~\bibnamefont {Engel}}, \bibinfo {author} {\bibfnamefont
  {C.}~\bibnamefont {Marcus}}, \bibinfo {author} {\bibfnamefont
  {D.}~\bibnamefont {Bono}}, \bibinfo {author} {\bibfnamefont {K.}~\bibnamefont
  {Bagschik}}, \bibinfo {author} {\bibfnamefont {S.}~\bibnamefont {Eisebitt}},\
  and\ \bibinfo {author} {\bibfnamefont {G.~S.~D.}\ \bibnamefont {Beach}},\
  }\bibfield  {title} {\bibinfo {title} {{Fast current-driven domain walls and
  small skyrmions in a compensated ferrimagnet}},\ }\href@noop {} {\bibfield
  {journal} {\bibinfo  {journal} {Nature Nanotechnology}\ } (\bibinfo {year}
  {2018})}\BibitemShut {NoStop}%
\bibitem [{\citenamefont {Hirata}\ \emph {et~al.}(2019)\citenamefont {Hirata},
  \citenamefont {Kim}, \citenamefont {Kim}, \citenamefont {Lee}, \citenamefont
  {Oh}, \citenamefont {Kim}, \citenamefont {Nishimura}, \citenamefont {Okuno},
  \citenamefont {Futakawa}, \citenamefont {Yoshikawa}, \citenamefont
  {Tsukamoto}, \citenamefont {Tserkovnyak}, \citenamefont {Shiota},
  \citenamefont {Moriyama}, \citenamefont {Choe}, \citenamefont {Lee},\ and\
  \citenamefont {Ono}}]{Hirata2019}%
  \BibitemOpen
  \bibfield  {author} {\bibinfo {author} {\bibfnamefont {Y.}~\bibnamefont
  {Hirata}}, \bibinfo {author} {\bibfnamefont {D.-H.}\ \bibnamefont {Kim}},
  \bibinfo {author} {\bibfnamefont {S.~K.}\ \bibnamefont {Kim}}, \bibinfo
  {author} {\bibfnamefont {D.-K.}\ \bibnamefont {Lee}}, \bibinfo {author}
  {\bibfnamefont {S.-H.}\ \bibnamefont {Oh}}, \bibinfo {author} {\bibfnamefont
  {D.-Y.}\ \bibnamefont {Kim}}, \bibinfo {author} {\bibfnamefont
  {T.}~\bibnamefont {Nishimura}}, \bibinfo {author} {\bibfnamefont
  {T.}~\bibnamefont {Okuno}}, \bibinfo {author} {\bibfnamefont
  {Y.}~\bibnamefont {Futakawa}}, \bibinfo {author} {\bibfnamefont
  {H.}~\bibnamefont {Yoshikawa}}, \bibinfo {author} {\bibfnamefont
  {A.}~\bibnamefont {Tsukamoto}}, \bibinfo {author} {\bibfnamefont
  {Y.}~\bibnamefont {Tserkovnyak}}, \bibinfo {author} {\bibfnamefont
  {Y.}~\bibnamefont {Shiota}}, \bibinfo {author} {\bibfnamefont
  {T.}~\bibnamefont {Moriyama}}, \bibinfo {author} {\bibfnamefont {S.-B.}\
  \bibnamefont {Choe}}, \bibinfo {author} {\bibfnamefont {K.-J.}\ \bibnamefont
  {Lee}},\ and\ \bibinfo {author} {\bibfnamefont {T.}~\bibnamefont {Ono}},\
  }\bibfield  {title} {\bibinfo {title} {{Vanishing skyrmion Hall effect at the
  angular momentum compensation temperature of a ferrimagnet}},\ }\href
  {https://doi.org/10.1038/s41565-018-0345-2} {\bibfield  {journal} {\bibinfo
  {journal} {Nature Nanotechnology}\ }\textbf {\bibinfo {volume} {14}},\
  \bibinfo {pages} {232} (\bibinfo {year} {2019})}\BibitemShut {NoStop}%
\bibitem [{\citenamefont {Woo}\ \emph {et~al.}(2018)\citenamefont {Woo},
  \citenamefont {Song}, \citenamefont {Zhang}, \citenamefont {Zhou},
  \citenamefont {Ezawa}, \citenamefont {Liu}, \citenamefont {Finizio},
  \citenamefont {Raabe}, \citenamefont {Lee}, \citenamefont {Kim},
  \citenamefont {Park}, \citenamefont {Kim}, \citenamefont {Kim}, \citenamefont
  {Lee}, \citenamefont {Lee}, \citenamefont {Choi}, \citenamefont {Min},
  \citenamefont {Koo},\ and\ \citenamefont {Chang}}]{Woo2018}%
  \BibitemOpen
  \bibfield  {author} {\bibinfo {author} {\bibfnamefont {S.}~\bibnamefont
  {Woo}}, \bibinfo {author} {\bibfnamefont {K.~M.}\ \bibnamefont {Song}},
  \bibinfo {author} {\bibfnamefont {X.}~\bibnamefont {Zhang}}, \bibinfo
  {author} {\bibfnamefont {Y.}~\bibnamefont {Zhou}}, \bibinfo {author}
  {\bibfnamefont {M.}~\bibnamefont {Ezawa}}, \bibinfo {author} {\bibfnamefont
  {X.}~\bibnamefont {Liu}}, \bibinfo {author} {\bibfnamefont {S.}~\bibnamefont
  {Finizio}}, \bibinfo {author} {\bibfnamefont {J.}~\bibnamefont {Raabe}},
  \bibinfo {author} {\bibfnamefont {N.~J.}\ \bibnamefont {Lee}}, \bibinfo
  {author} {\bibfnamefont {S.-I.}\ \bibnamefont {Kim}}, \bibinfo {author}
  {\bibfnamefont {S.-Y.}\ \bibnamefont {Park}}, \bibinfo {author}
  {\bibfnamefont {Y.}~\bibnamefont {Kim}}, \bibinfo {author} {\bibfnamefont
  {J.-Y.}\ \bibnamefont {Kim}}, \bibinfo {author} {\bibfnamefont
  {D.}~\bibnamefont {Lee}}, \bibinfo {author} {\bibfnamefont {O.}~\bibnamefont
  {Lee}}, \bibinfo {author} {\bibfnamefont {J.~W.}\ \bibnamefont {Choi}},
  \bibinfo {author} {\bibfnamefont {B.-C.}\ \bibnamefont {Min}}, \bibinfo
  {author} {\bibfnamefont {H.~C.}\ \bibnamefont {Koo}},\ and\ \bibinfo {author}
  {\bibfnamefont {J.}~\bibnamefont {Chang}},\ }\bibfield  {title} {\bibinfo
  {title} {{Current-driven dynamics and inhibition of the skyrmion Hall effect
  of ferrimagnetic skyrmions in GdFeCo films}},\ }\href
  {https://doi.org/10.1038/s41467-018-03378-7} {\bibfield  {journal} {\bibinfo
  {journal} {Nature Communications}\ }\textbf {\bibinfo {volume} {9}},\
  \bibinfo {pages} {959} (\bibinfo {year} {2018})}\BibitemShut {NoStop}%
\bibitem [{\citenamefont {Haltz}\ \emph {et~al.}(2019)\citenamefont {Haltz},
  \citenamefont {Sampaio}, \citenamefont {Weil}, \citenamefont {Dumont},\ and\
  \citenamefont {Mougin}}]{Haltz2019}%
  \BibitemOpen
  \bibfield  {author} {\bibinfo {author} {\bibfnamefont {E.}~\bibnamefont
  {Haltz}}, \bibinfo {author} {\bibfnamefont {J.}~\bibnamefont {Sampaio}},
  \bibinfo {author} {\bibfnamefont {R.}~\bibnamefont {Weil}}, \bibinfo {author}
  {\bibfnamefont {Y.}~\bibnamefont {Dumont}},\ and\ \bibinfo {author}
  {\bibfnamefont {A.}~\bibnamefont {Mougin}},\ }\bibfield  {title} {\bibinfo
  {title} {{Strong current actions on ferrimagnetic domain walls in the creep
  regime}},\ }\href {https://doi.org/10.1103/PhysRevB.99.104413} {\bibfield
  {journal} {\bibinfo  {journal} {Physical Review B}\ }\textbf {\bibinfo
  {volume} {99}},\ \bibinfo {pages} {104413} (\bibinfo {year}
  {2019})}\BibitemShut {NoStop}%
\bibitem [{\citenamefont {Haltz}\ \emph {et~al.}(2020)\citenamefont {Haltz},
  \citenamefont {Sampaio}, \citenamefont {Krishnia}, \citenamefont {Berges},
  \citenamefont {Weil},\ and\ \citenamefont {Mougin}}]{Haltz2020}%
  \BibitemOpen
  \bibfield  {author} {\bibinfo {author} {\bibfnamefont {E.}~\bibnamefont
  {Haltz}}, \bibinfo {author} {\bibfnamefont {J.}~\bibnamefont {Sampaio}},
  \bibinfo {author} {\bibfnamefont {S.}~\bibnamefont {Krishnia}}, \bibinfo
  {author} {\bibfnamefont {L.}~\bibnamefont {Berges}}, \bibinfo {author}
  {\bibfnamefont {R.}~\bibnamefont {Weil}},\ and\ \bibinfo {author}
  {\bibfnamefont {A.}~\bibnamefont {Mougin}},\ }\bibfield  {title} {\bibinfo
  {title} {{Measurement of the tilt of a moving domain wall shows
  precession-free dynamics in compensated ferrimagnets}},\ }\href
  {https://doi.org/10.1038/s41598-020-73049-5} {\bibfield  {journal} {\bibinfo
  {journal} {Scientific Reports}\ }\textbf {\bibinfo {volume} {10}},\ \bibinfo
  {pages} {16292} (\bibinfo {year} {2020})}\BibitemShut {NoStop}%
\bibitem [{\citenamefont {Shiino}\ \emph {et~al.}(2016)\citenamefont {Shiino},
  \citenamefont {Oh}, \citenamefont {Haney}, \citenamefont {Lee}, \citenamefont
  {Go}, \citenamefont {Park},\ and\ \citenamefont {Lee}}]{Shiino2016}%
  \BibitemOpen
  \bibfield  {author} {\bibinfo {author} {\bibfnamefont {T.}~\bibnamefont
  {Shiino}}, \bibinfo {author} {\bibfnamefont {S.-H.}\ \bibnamefont {Oh}},
  \bibinfo {author} {\bibfnamefont {P.~M.}\ \bibnamefont {Haney}}, \bibinfo
  {author} {\bibfnamefont {S.-W.}\ \bibnamefont {Lee}}, \bibinfo {author}
  {\bibfnamefont {G.}~\bibnamefont {Go}}, \bibinfo {author} {\bibfnamefont
  {B.-G.}\ \bibnamefont {Park}},\ and\ \bibinfo {author} {\bibfnamefont
  {K.-J.}\ \bibnamefont {Lee}},\ }\bibfield  {title} {\bibinfo {title}
  {{Antiferromagnetic Domain Wall Motion Driven by Spin-Orbit Torques}},\
  }\href {https://doi.org/10.1103/PhysRevLett.117.087203} {\bibfield  {journal}
  {\bibinfo  {journal} {Physical Review Letters}\ }\textbf {\bibinfo {volume}
  {117}},\ \bibinfo {pages} {087203} (\bibinfo {year} {2016})}\BibitemShut
  {NoStop}%
\bibitem [{\citenamefont {Keffer}\ and\ \citenamefont
  {Kittel}(1952)}]{Keffer1952}%
  \BibitemOpen
  \bibfield  {author} {\bibinfo {author} {\bibfnamefont {F.}~\bibnamefont
  {Keffer}}\ and\ \bibinfo {author} {\bibfnamefont {C.}~\bibnamefont
  {Kittel}},\ }\bibfield  {title} {\bibinfo {title} {{Theory of
  antiferromagnetic resonance}},\ }\href
  {https://doi.org/10.1103/PhysRev.85.329} {\bibfield  {journal} {\bibinfo
  {journal} {Phys. Rev.}\ }\textbf {\bibinfo {volume} {85}},\ \bibinfo {pages}
  {329} (\bibinfo {year} {1952})}\BibitemShut {NoStop}%
\bibitem [{\citenamefont {Stremoukhov}\ \emph {et~al.}(2019)\citenamefont
  {Stremoukhov}, \citenamefont {Safin}, \citenamefont {Logunov}, \citenamefont
  {Nikitov},\ and\ \citenamefont {Kirilyuk}}]{Stremoukhov2019}%
  \BibitemOpen
  \bibfield  {author} {\bibinfo {author} {\bibfnamefont {P.}~\bibnamefont
  {Stremoukhov}}, \bibinfo {author} {\bibfnamefont {A.}~\bibnamefont {Safin}},
  \bibinfo {author} {\bibfnamefont {M.}~\bibnamefont {Logunov}}, \bibinfo
  {author} {\bibfnamefont {S.}~\bibnamefont {Nikitov}},\ and\ \bibinfo {author}
  {\bibfnamefont {A.}~\bibnamefont {Kirilyuk}},\ }\bibfield  {title} {\bibinfo
  {title} {{Spintronic terahertz-frequency nonlinear emitter based on the
  canted antiferromagnet-platinum bilayers}},\ }\href
  {https://doi.org/10.1063/1.5090455} {\bibfield  {journal} {\bibinfo
  {journal} {J. Appl. Phys.}\ }\textbf {\bibinfo {volume} {125}},\ \bibinfo
  {pages} {223903} (\bibinfo {year} {2019})}\BibitemShut {NoStop}%
\bibitem [{\citenamefont {Oh}\ \emph {et~al.}(2017)\citenamefont {Oh},
  \citenamefont {Kim}, \citenamefont {Lee}, \citenamefont {Go}, \citenamefont
  {Kim}, \citenamefont {Ono}, \citenamefont {Tserkovnyak},\ and\ \citenamefont
  {Lee}}]{Oh2017}%
  \BibitemOpen
  \bibfield  {author} {\bibinfo {author} {\bibfnamefont {S.-H.}\ \bibnamefont
  {Oh}}, \bibinfo {author} {\bibfnamefont {S.~K.}\ \bibnamefont {Kim}},
  \bibinfo {author} {\bibfnamefont {D.-K.}\ \bibnamefont {Lee}}, \bibinfo
  {author} {\bibfnamefont {G.}~\bibnamefont {Go}}, \bibinfo {author}
  {\bibfnamefont {K.-J.}\ \bibnamefont {Kim}}, \bibinfo {author} {\bibfnamefont
  {T.}~\bibnamefont {Ono}}, \bibinfo {author} {\bibfnamefont {Y.}~\bibnamefont
  {Tserkovnyak}},\ and\ \bibinfo {author} {\bibfnamefont {K.-J.}\ \bibnamefont
  {Lee}},\ }\bibfield  {title} {\bibinfo {title} {{Coherent terahertz spin-wave
  emission associated with ferrimagnetic domain wall dynamics}},\ }\href
  {https://doi.org/10.1103/PhysRevB.96.100407} {\bibfield  {journal} {\bibinfo
  {journal} {Physical Review B}\ }\textbf {\bibinfo {volume} {96}},\ \bibinfo
  {pages} {100407} (\bibinfo {year} {2017})}\BibitemShut {NoStop}%
\bibitem [{\citenamefont {Lepadatu}\ \emph {et~al.}(2017)\citenamefont
  {Lepadatu}, \citenamefont {Saarikoski}, \citenamefont {Beacham},
  \citenamefont {Benitez}, \citenamefont {Moore}, \citenamefont {Burnell},
  \citenamefont {Sugimoto}, \citenamefont {Yesudas}, \citenamefont {Wheeler},
  \citenamefont {Miguel}, \citenamefont {Dhesi}, \citenamefont {McGrouther},
  \citenamefont {McVitie}, \citenamefont {Tatara},\ and\ \citenamefont
  {Marrows}}]{Lepadatu2017}%
  \BibitemOpen
  \bibfield  {author} {\bibinfo {author} {\bibfnamefont {S.}~\bibnamefont
  {Lepadatu}}, \bibinfo {author} {\bibfnamefont {H.}~\bibnamefont
  {Saarikoski}}, \bibinfo {author} {\bibfnamefont {R.}~\bibnamefont {Beacham}},
  \bibinfo {author} {\bibfnamefont {M.~J.}\ \bibnamefont {Benitez}}, \bibinfo
  {author} {\bibfnamefont {T.~A.}\ \bibnamefont {Moore}}, \bibinfo {author}
  {\bibfnamefont {G.}~\bibnamefont {Burnell}}, \bibinfo {author} {\bibfnamefont
  {S.}~\bibnamefont {Sugimoto}}, \bibinfo {author} {\bibfnamefont
  {D.}~\bibnamefont {Yesudas}}, \bibinfo {author} {\bibfnamefont {M.~C.}\
  \bibnamefont {Wheeler}}, \bibinfo {author} {\bibfnamefont {J.}~\bibnamefont
  {Miguel}}, \bibinfo {author} {\bibfnamefont {S.~S.}\ \bibnamefont {Dhesi}},
  \bibinfo {author} {\bibfnamefont {D.}~\bibnamefont {McGrouther}}, \bibinfo
  {author} {\bibfnamefont {S.}~\bibnamefont {McVitie}}, \bibinfo {author}
  {\bibfnamefont {G.}~\bibnamefont {Tatara}},\ and\ \bibinfo {author}
  {\bibfnamefont {C.~H.}\ \bibnamefont {Marrows}},\ }\bibfield  {title}
  {\bibinfo {title} {{Synthetic ferrimagnet nanowires with very low critical
  current density for coupled domain wall motion}},\ }\href
  {https://doi.org/10.1038/s41598-017-01748-7} {\bibfield  {journal} {\bibinfo
  {journal} {Scientific Reports}\ }\textbf {\bibinfo {volume} {7}},\ \bibinfo
  {pages} {1640} (\bibinfo {year} {2017})}\BibitemShut {NoStop}%
\bibitem [{\citenamefont {Duine}\ \emph {et~al.}(2018)\citenamefont {Duine},
  \citenamefont {Lee}, \citenamefont {Parkin},\ and\ \citenamefont
  {Stiles}}]{Duine2018}%
  \BibitemOpen
  \bibfield  {author} {\bibinfo {author} {\bibfnamefont {R.~A.}\ \bibnamefont
  {Duine}}, \bibinfo {author} {\bibfnamefont {K.-J.}\ \bibnamefont {Lee}},
  \bibinfo {author} {\bibfnamefont {S.~S.~P.}\ \bibnamefont {Parkin}},\ and\
  \bibinfo {author} {\bibfnamefont {M.~D.}\ \bibnamefont {Stiles}},\ }\bibfield
   {title} {\bibinfo {title} {{Synthetic antiferromagnetic spintronics}},\
  }\href {https://doi.org/10.1038/s41567-018-0050-y} {\bibfield  {journal}
  {\bibinfo  {journal} {Nature Physics}\ }\textbf {\bibinfo {volume} {14}},\
  \bibinfo {pages} {217} (\bibinfo {year} {2018})}\BibitemShut {NoStop}%
\bibitem [{\citenamefont {Parkin}\ \emph {et~al.}(1990)\citenamefont {Parkin},
  \citenamefont {More},\ and\ \citenamefont {Roche}}]{Parkin1990}%
  \BibitemOpen
  \bibfield  {author} {\bibinfo {author} {\bibfnamefont {S.~S.~P.}\
  \bibnamefont {Parkin}}, \bibinfo {author} {\bibfnamefont {N.}~\bibnamefont
  {More}},\ and\ \bibinfo {author} {\bibfnamefont {K.~P.}\ \bibnamefont
  {Roche}},\ }\bibfield  {title} {\bibinfo {title} {{Oscillations in exchange
  coupling and magnetoresistance in metallic superlattice structures: Co/Ru,
  Co/Cr, and Fe/Cr}},\ }\href {https://doi.org/10.1103/PhysRevLett.64.2304}
  {\bibfield  {journal} {\bibinfo  {journal} {Physical Review Letters}\
  }\textbf {\bibinfo {volume} {64}},\ \bibinfo {pages} {2304} (\bibinfo {year}
  {1990})}\BibitemShut {NoStop}%
\bibitem [{\citenamefont {Parkin}(1991)}]{Parkin1991}%
  \BibitemOpen
  \bibfield  {author} {\bibinfo {author} {\bibfnamefont {S.~S.~P.}\
  \bibnamefont {Parkin}},\ }\bibfield  {title} {\bibinfo {title} {{Systematic
  variation of the strength and oscillation period of indirect magnetic
  exchange coupling through the 3d, 4d, and 5d transition metals}},\ }\href
  {https://doi.org/10.1103/PhysRevLett.67.3598} {\bibfield  {journal} {\bibinfo
   {journal} {Phys. Rev. Lett.}\ }\textbf {\bibinfo {volume} {67}},\ \bibinfo
  {pages} {3598} (\bibinfo {year} {1991})}\BibitemShut {NoStop}%
\bibitem [{\citenamefont {Hansen}\ \emph {et~al.}(1989)\citenamefont {Hansen},
  \citenamefont {Clausen}, \citenamefont {Much}, \citenamefont {Rosenkranz},\
  and\ \citenamefont {Witter}}]{Hansen1989}%
  \BibitemOpen
  \bibfield  {author} {\bibinfo {author} {\bibfnamefont {P.}~\bibnamefont
  {Hansen}}, \bibinfo {author} {\bibfnamefont {C.}~\bibnamefont {Clausen}},
  \bibinfo {author} {\bibfnamefont {G.}~\bibnamefont {Much}}, \bibinfo {author}
  {\bibfnamefont {M.}~\bibnamefont {Rosenkranz}},\ and\ \bibinfo {author}
  {\bibfnamefont {K.}~\bibnamefont {Witter}},\ }\bibfield  {title} {\bibinfo
  {title} {{Magnetic and magneto-optical properties of rare-earth
  transition-metal alloys containing Gd, Tb, Fe, Co}},\ }\href
  {https://doi.org/10.1063/1.343551} {\bibfield  {journal} {\bibinfo  {journal}
  {Journal of Applied Physics}\ }\textbf {\bibinfo {volume} {66}},\ \bibinfo
  {pages} {756} (\bibinfo {year} {1989})}\BibitemShut {NoStop}%
\bibitem [{\citenamefont {Hirata}\ \emph {et~al.}(2018)\citenamefont {Hirata},
  \citenamefont {Kim}, \citenamefont {Okuno}, \citenamefont {Nishimura},
  \citenamefont {Kim}, \citenamefont {Futakawa}, \citenamefont {Yoshikawa},
  \citenamefont {Tsukamoto}, \citenamefont {Kim}, \citenamefont {Choe},\ and\
  \citenamefont {Ono}}]{Hirata2018}%
  \BibitemOpen
  \bibfield  {author} {\bibinfo {author} {\bibfnamefont {Y.}~\bibnamefont
  {Hirata}}, \bibinfo {author} {\bibfnamefont {D.-H.}\ \bibnamefont {Kim}},
  \bibinfo {author} {\bibfnamefont {T.}~\bibnamefont {Okuno}}, \bibinfo
  {author} {\bibfnamefont {T.}~\bibnamefont {Nishimura}}, \bibinfo {author}
  {\bibfnamefont {D.-Y.}\ \bibnamefont {Kim}}, \bibinfo {author} {\bibfnamefont
  {Y.}~\bibnamefont {Futakawa}}, \bibinfo {author} {\bibfnamefont
  {H.}~\bibnamefont {Yoshikawa}}, \bibinfo {author} {\bibfnamefont
  {A.}~\bibnamefont {Tsukamoto}}, \bibinfo {author} {\bibfnamefont {K.-J.}\
  \bibnamefont {Kim}}, \bibinfo {author} {\bibfnamefont {S.-B.}\ \bibnamefont
  {Choe}},\ and\ \bibinfo {author} {\bibfnamefont {T.}~\bibnamefont {Ono}},\
  }\bibfield  {title} {\bibinfo {title} {{Correlation between compensation
  temperatures of magnetization and angular momentum in GdFeCo ferrimagnets}},\
  }\href {https://doi.org/10.1103/PhysRevB.97.220403} {\bibfield  {journal}
  {\bibinfo  {journal} {Physical Review B}\ }\textbf {\bibinfo {volume} {97}},\
  \bibinfo {pages} {220403} (\bibinfo {year} {2018})}\BibitemShut {NoStop}%
\bibitem [{\citenamefont {Hagedorn}(1972)}]{Hagedorn1972}%
  \BibitemOpen
  \bibfield  {author} {\bibinfo {author} {\bibfnamefont {F.~B.}\ \bibnamefont
  {Hagedorn}},\ }\bibfield  {title} {\bibinfo {title} {{Domain wall motion in
  bubble domain materials}},\ }\href {https://doi.org/10.1063/1.3699399}
  {\bibfield  {journal} {\bibinfo  {journal} {AIP Conference Proceedings}\
  }\textbf {\bibinfo {volume} {5}},\ \bibinfo {pages} {72} (\bibinfo {year}
  {1972})}\BibitemShut {NoStop}%
\bibitem [{\citenamefont {Landau}\ and\ \citenamefont
  {Lifshitz}(1935)}]{Landau1935}%
  \BibitemOpen
  \bibfield  {author} {\bibinfo {author} {\bibfnamefont {L.~D.}\ \bibnamefont
  {Landau}}\ and\ \bibinfo {author} {\bibfnamefont {E.}~\bibnamefont
  {Lifshitz}},\ }\bibfield  {title} {\bibinfo {title} {{On the theory of the
  dispersion of magnetic permeability in ferromagnetic bodies}},\ }\href@noop
  {} {\bibfield  {journal} {\bibinfo  {journal} {Physik Zeitsch. der
  Sowjetunion}\ }\textbf {\bibinfo {volume} {8}},\ \bibinfo {pages} {153}
  (\bibinfo {year} {1935})}\BibitemShut {NoStop}%
\bibitem [{\citenamefont {Gilbert}(2004)}]{Gilbert2004}%
  \BibitemOpen
  \bibfield  {author} {\bibinfo {author} {\bibfnamefont {T.}~\bibnamefont
  {Gilbert}},\ }\bibfield  {title} {\bibinfo {title} {{A Phenomenological
  Theory of Damping in Ferromagnetic Materials}},\ }\href
  {https://doi.org/10.1109/TMAG.2004.836740} {\bibfield  {journal} {\bibinfo
  {journal} {IEEE Transactions on Magnetics}\ }\textbf {\bibinfo {volume}
  {40}},\ \bibinfo {pages} {3443} (\bibinfo {year} {2004})}\BibitemShut
  {NoStop}%
\bibitem [{Note1()}]{Note1}%
  \BibitemOpen
  \bibinfo {note} {If $\protect \vec {\protect \bm {\tau }}\protect \neq 0$,
  the energy variation is instead $d_tU= -L_\alpha \left | \partial _t \protect
  \vec {\protect \bm {m}} \right |^2 + \protect \frac {1}{L_S}\protect \vec
  {\protect \bm {\tau }}\cdot \left ( L_\alpha \partial _t \protect \vec
  {\protect \bm {m}}+ \delta _{\protect \vec {\protect \bm {m}}}U\right )$, or
  $d_tU=-\protect \frac {L_\alpha }{L_S^2+L_\alpha ^2} \left ( |\protect \vec
  {\protect \bm {m}}\times \delta _{\protect \vec {\protect \bm {m}}}U|^2
  -\protect \vec {\protect \bm {\tau }}\cdot (\protect \vec {\protect \bm
  {m}}\times \delta _{\protect \vec {\protect \bm {m}}}U) \right )+\protect
  \frac {L_S}{L_S^2+L_\alpha ^2} \delta _{\protect \vec {\protect \bm
  {m}}}U\cdot \protect \vec {\protect \bm {\tau }}$}\BibitemShut {NoStop}%
\bibitem [{\citenamefont {Thiaville}\ \emph {et~al.}(2005)\citenamefont
  {Thiaville}, \citenamefont {Nakatani}, \citenamefont {Miltat},\ and\
  \citenamefont {Suzuki}}]{Thiaville2005}%
  \BibitemOpen
  \bibfield  {author} {\bibinfo {author} {\bibfnamefont {A.}~\bibnamefont
  {Thiaville}}, \bibinfo {author} {\bibfnamefont {Y.}~\bibnamefont {Nakatani}},
  \bibinfo {author} {\bibfnamefont {J.}~\bibnamefont {Miltat}},\ and\ \bibinfo
  {author} {\bibfnamefont {Y.}~\bibnamefont {Suzuki}},\ }\bibfield  {title}
  {\bibinfo {title} {{Micromagnetic understanding of current-driven domain wall
  motion in patterned nanowires}},\ }\href
  {https://doi.org/10.1209/epl/i2004-10452-6} {\bibfield  {journal} {\bibinfo
  {journal} {Europhysics Letters}\ }\textbf {\bibinfo {volume} {69}},\ \bibinfo
  {pages} {990} (\bibinfo {year} {2005})}\BibitemShut {NoStop}%
\bibitem [{\citenamefont {Gomonay}\ and\ \citenamefont
  {Loktev}(2014)}]{Gomonay2014}%
  \BibitemOpen
  \bibfield  {author} {\bibinfo {author} {\bibfnamefont {E.~V.}\ \bibnamefont
  {Gomonay}}\ and\ \bibinfo {author} {\bibfnamefont {V.~M.}\ \bibnamefont
  {Loktev}},\ }\bibfield  {title} {\bibinfo {title} {{Spintronics of
  antiferromagnetic systems (Review Article)}},\ }\href
  {https://doi.org/10.1063/1.4862467} {\bibfield  {journal} {\bibinfo
  {journal} {Low Temperature Physics}\ }\textbf {\bibinfo {volume} {40}},\
  \bibinfo {pages} {17} (\bibinfo {year} {2014})}\BibitemShut {NoStop}%
\bibitem [{\citenamefont {Wangsness}(1953)}]{Wangsness1953}%
  \BibitemOpen
  \bibfield  {author} {\bibinfo {author} {\bibfnamefont {R.~K.}\ \bibnamefont
  {Wangsness}},\ }\bibfield  {title} {\bibinfo {title} {{Sublattice Effects in
  Magnetic Resonance}},\ }\href
  {https://link.aps.org/doi/10.1103/PhysRev.91.1085} {\bibfield  {journal}
  {\bibinfo  {journal} {Physical Review}\ }\textbf {\bibinfo {volume} {91}},\
  \bibinfo {pages} {1085} (\bibinfo {year} {1953})}\BibitemShut {NoStop}%
\bibitem [{\citenamefont {Bl{\"{a}}sing}\ \emph {et~al.}(2018)\citenamefont
  {Bl{\"{a}}sing}, \citenamefont {Ma}, \citenamefont {Yang}, \citenamefont
  {Garg}, \citenamefont {Dejene}, \citenamefont {N'Diaye}, \citenamefont
  {Chen}, \citenamefont {Liu},\ and\ \citenamefont {Parkin}}]{Blasing2018}%
  \BibitemOpen
  \bibfield  {author} {\bibinfo {author} {\bibfnamefont {R.}~\bibnamefont
  {Bl{\"{a}}sing}}, \bibinfo {author} {\bibfnamefont {T.}~\bibnamefont {Ma}},
  \bibinfo {author} {\bibfnamefont {S.-H.}\ \bibnamefont {Yang}}, \bibinfo
  {author} {\bibfnamefont {C.}~\bibnamefont {Garg}}, \bibinfo {author}
  {\bibfnamefont {F.~K.}\ \bibnamefont {Dejene}}, \bibinfo {author}
  {\bibfnamefont {A.~T.}\ \bibnamefont {N'Diaye}}, \bibinfo {author}
  {\bibfnamefont {G.}~\bibnamefont {Chen}}, \bibinfo {author} {\bibfnamefont
  {K.}~\bibnamefont {Liu}},\ and\ \bibinfo {author} {\bibfnamefont {S.~S.~P.}\
  \bibnamefont {Parkin}},\ }\bibfield  {title} {\bibinfo {title} {{Exchange
  coupling torque in ferrimagnetic Co/Gd bilayer maximized near angular
  momentum compensation temperature}},\ }\href
  {https://doi.org/10.1038/s41467-018-07373-w} {\bibfield  {journal} {\bibinfo
  {journal} {Nature Communications}\ }\textbf {\bibinfo {volume} {9}},\
  \bibinfo {pages} {4984} (\bibinfo {year} {2018})}\BibitemShut {NoStop}%
\bibitem [{\citenamefont {Kim}\ \emph {et~al.}(2019)\citenamefont {Kim},
  \citenamefont {Okuno}, \citenamefont {Kim}, \citenamefont {Oh}, \citenamefont
  {Nishimura}, \citenamefont {Hirata}, \citenamefont {Futakawa}, \citenamefont
  {Yoshikawa}, \citenamefont {Tsukamoto}, \citenamefont {Tserkovnyak},
  \citenamefont {Shiota}, \citenamefont {Moriyama}, \citenamefont {Kim},
  \citenamefont {Lee},\ and\ \citenamefont {Ono}}]{Kim2019}%
  \BibitemOpen
  \bibfield  {author} {\bibinfo {author} {\bibfnamefont {D.-H.}\ \bibnamefont
  {Kim}}, \bibinfo {author} {\bibfnamefont {T.}~\bibnamefont {Okuno}}, \bibinfo
  {author} {\bibfnamefont {S.~K.}\ \bibnamefont {Kim}}, \bibinfo {author}
  {\bibfnamefont {S.-H.}\ \bibnamefont {Oh}}, \bibinfo {author} {\bibfnamefont
  {T.}~\bibnamefont {Nishimura}}, \bibinfo {author} {\bibfnamefont
  {Y.}~\bibnamefont {Hirata}}, \bibinfo {author} {\bibfnamefont
  {Y.}~\bibnamefont {Futakawa}}, \bibinfo {author} {\bibfnamefont
  {H.}~\bibnamefont {Yoshikawa}}, \bibinfo {author} {\bibfnamefont
  {A.}~\bibnamefont {Tsukamoto}}, \bibinfo {author} {\bibfnamefont
  {Y.}~\bibnamefont {Tserkovnyak}}, \bibinfo {author} {\bibfnamefont
  {Y.}~\bibnamefont {Shiota}}, \bibinfo {author} {\bibfnamefont
  {T.}~\bibnamefont {Moriyama}}, \bibinfo {author} {\bibfnamefont {K.-J.}\
  \bibnamefont {Kim}}, \bibinfo {author} {\bibfnamefont {K.-J.}\ \bibnamefont
  {Lee}},\ and\ \bibinfo {author} {\bibfnamefont {T.}~\bibnamefont {Ono}},\
  }\bibfield  {title} {\bibinfo {title} {{Low Magnetic Damping of Ferrimagnetic
  GdFeCo Alloys}},\ }\href {https://doi.org/10.1103/PhysRevLett.122.127203}
  {\bibfield  {journal} {\bibinfo  {journal} {Physical Review Letters}\
  }\textbf {\bibinfo {volume} {122}},\ \bibinfo {pages} {127203} (\bibinfo
  {year} {2019})}\BibitemShut {NoStop}%
\bibitem [{\citenamefont {Krishnia}\ \emph {et~al.}(2020)\citenamefont
  {Krishnia}, \citenamefont {Haltz}, \citenamefont {Berges}, \citenamefont
  {Aballe}, \citenamefont {Foerster}, \citenamefont {Bocher}, \citenamefont
  {Weil}, \citenamefont {Thiaville}, \citenamefont {Sampaio},\ and\
  \citenamefont {Mougin}}]{Krishnia2020}%
  \BibitemOpen
  \bibfield  {author} {\bibinfo {author} {\bibfnamefont {S.}~\bibnamefont
  {Krishnia}}, \bibinfo {author} {\bibfnamefont {E.}~\bibnamefont {Haltz}},
  \bibinfo {author} {\bibfnamefont {L.}~\bibnamefont {Berges}}, \bibinfo
  {author} {\bibfnamefont {L.}~\bibnamefont {Aballe}}, \bibinfo {author}
  {\bibfnamefont {M.}~\bibnamefont {Foerster}}, \bibinfo {author}
  {\bibfnamefont {L.}~\bibnamefont {Bocher}}, \bibinfo {author} {\bibfnamefont
  {R.}~\bibnamefont {Weil}}, \bibinfo {author} {\bibfnamefont {A.}~\bibnamefont
  {Thiaville}}, \bibinfo {author} {\bibfnamefont {J.}~\bibnamefont {Sampaio}},\
  and\ \bibinfo {author} {\bibfnamefont {A.}~\bibnamefont {Mougin}},\
  }\bibfield  {title} {\bibinfo {title} {{Direct observation of chiral domain
  walls and spin-orbit torques in single-layer ferrimagnetic alloys}},\
  }\href@noop {} {\bibfield  {journal} {\bibinfo  {journal} {submitted.}\ }
  (\bibinfo {year} {2020})}\BibitemShut {NoStop}%
\bibitem [{\citenamefont {Schryer}\ and\ \citenamefont
  {Walker}(1974)}]{Schryer1974}%
  \BibitemOpen
  \bibfield  {author} {\bibinfo {author} {\bibfnamefont {N.~L.}\ \bibnamefont
  {Schryer}}\ and\ \bibinfo {author} {\bibfnamefont {L.~R.}\ \bibnamefont
  {Walker}},\ }\bibfield  {title} {\bibinfo {title} {{The motion of 180 domain
  walls in uniform dc magnetic fields}},\ }\href
  {https://doi.org/10.1063/1.1663252} {\bibfield  {journal} {\bibinfo
  {journal} {Journal of Applied Physics}\ }\textbf {\bibinfo {volume} {45}},\
  \bibinfo {pages} {5406} (\bibinfo {year} {1974})}\BibitemShut {NoStop}%
\bibitem [{\citenamefont {Mougin}\ \emph {et~al.}(2007)\citenamefont {Mougin},
  \citenamefont {Cormier}, \citenamefont {Adam}, \citenamefont {Metaxas},\ and\
  \citenamefont {Ferr{\'{e}}}}]{Mougin2007}%
  \BibitemOpen
  \bibfield  {author} {\bibinfo {author} {\bibfnamefont {A.}~\bibnamefont
  {Mougin}}, \bibinfo {author} {\bibfnamefont {M.}~\bibnamefont {Cormier}},
  \bibinfo {author} {\bibfnamefont {J.}~\bibnamefont {Adam}}, \bibinfo {author}
  {\bibfnamefont {P.~J.}\ \bibnamefont {Metaxas}},\ and\ \bibinfo {author}
  {\bibfnamefont {J.}~\bibnamefont {Ferr{\'{e}}}},\ }\bibfield  {title}
  {\bibinfo {title} {{Domain wall mobility, stability and Walker breakdown in
  magnetic nanowires}},\ }\href {https://doi.org/10.1209/0295-5075/78/57007}
  {\bibfield  {journal} {\bibinfo  {journal} {Europhysics Letters}\ }\textbf
  {\bibinfo {volume} {78}},\ \bibinfo {pages} {57007} (\bibinfo {year}
  {2007})}\BibitemShut {NoStop}%
\bibitem [{\citenamefont {Thiaville}\ \emph {et~al.}(2012)\citenamefont
  {Thiaville}, \citenamefont {Rohart}, \citenamefont {Ju{\'{e}}}, \citenamefont
  {Cros},\ and\ \citenamefont {Fert}}]{Thiaville2012}%
  \BibitemOpen
  \bibfield  {author} {\bibinfo {author} {\bibfnamefont {A.}~\bibnamefont
  {Thiaville}}, \bibinfo {author} {\bibfnamefont {S.}~\bibnamefont {Rohart}},
  \bibinfo {author} {\bibfnamefont {E.}~\bibnamefont {Ju{\'{e}}}}, \bibinfo
  {author} {\bibfnamefont {V.}~\bibnamefont {Cros}},\ and\ \bibinfo {author}
  {\bibfnamefont {A.}~\bibnamefont {Fert}},\ }\bibfield  {title} {\bibinfo
  {title} {{Dynamics of Dzyaloshinskii domain walls in ultrathin magnetic
  films}},\ }\href {https://doi.org/10.1209/0295-5075/100/57002} {\bibfield
  {journal} {\bibinfo  {journal} {EPL}\ }\textbf {\bibinfo {volume} {100}},\
  \bibinfo {pages} {57002} (\bibinfo {year} {2012})}\BibitemShut {NoStop}%
\bibitem [{\citenamefont {Malozemoff}\ and\ \citenamefont
  {Slonczewski}(1979)}]{Malozemoff1979}%
  \BibitemOpen
  \bibfield  {author} {\bibinfo {author} {\bibfnamefont {A.}~\bibnamefont
  {Malozemoff}}\ and\ \bibinfo {author} {\bibfnamefont {J.}~\bibnamefont
  {Slonczewski}},\ }\href {https://doi.org/10.1016/C2013-0-06998-8} {\emph
  {\bibinfo {title} {Magnetic Domain Walls in Bubble Materials}}}\ (\bibinfo
  {publisher} {Academic Press},\ \bibinfo {year} {1979})\ pp.\ \bibinfo {pages}
  {269--292}\BibitemShut {NoStop}%
\bibitem [{\citenamefont {Vansteenkiste}\ \emph {et~al.}(2014)\citenamefont
  {Vansteenkiste}, \citenamefont {Leliaert}, \citenamefont {Dvornik},
  \citenamefont {Helsen}, \citenamefont {Garcia-Sanchez},\ and\ \citenamefont
  {{Van Waeyenberge}}}]{Vansteenkiste2014}%
  \BibitemOpen
  \bibfield  {author} {\bibinfo {author} {\bibfnamefont {A.}~\bibnamefont
  {Vansteenkiste}}, \bibinfo {author} {\bibfnamefont {J.}~\bibnamefont
  {Leliaert}}, \bibinfo {author} {\bibfnamefont {M.}~\bibnamefont {Dvornik}},
  \bibinfo {author} {\bibfnamefont {M.}~\bibnamefont {Helsen}}, \bibinfo
  {author} {\bibfnamefont {F.}~\bibnamefont {Garcia-Sanchez}},\ and\ \bibinfo
  {author} {\bibfnamefont {B.}~\bibnamefont {{Van Waeyenberge}}},\ }\bibfield
  {title} {\bibinfo {title} {{The design and verification of MuMax3}},\ }\href
  {https://doi.org/10.1063/1.4899186} {\bibfield  {journal} {\bibinfo
  {journal} {AIP Advances}\ }\textbf {\bibinfo {volume} {4}},\ \bibinfo {pages}
  {107133} (\bibinfo {year} {2014})}\BibitemShut {NoStop}%
\bibitem [{\citenamefont {Haltz}\ \emph {et~al.}(2018)\citenamefont {Haltz},
  \citenamefont {Weil}, \citenamefont {Sampaio}, \citenamefont {Pointillon},
  \citenamefont {Rousseau}, \citenamefont {March}, \citenamefont {Brun},
  \citenamefont {Li}, \citenamefont {Briand}, \citenamefont {Bachelet},
  \citenamefont {Dumont},\ and\ \citenamefont {Mougin}}]{Haltz2018}%
  \BibitemOpen
  \bibfield  {author} {\bibinfo {author} {\bibfnamefont {E.}~\bibnamefont
  {Haltz}}, \bibinfo {author} {\bibfnamefont {R.}~\bibnamefont {Weil}},
  \bibinfo {author} {\bibfnamefont {J.}~\bibnamefont {Sampaio}}, \bibinfo
  {author} {\bibfnamefont {A.}~\bibnamefont {Pointillon}}, \bibinfo {author}
  {\bibfnamefont {O.}~\bibnamefont {Rousseau}}, \bibinfo {author}
  {\bibfnamefont {K.}~\bibnamefont {March}}, \bibinfo {author} {\bibfnamefont
  {N.}~\bibnamefont {Brun}}, \bibinfo {author} {\bibfnamefont {Z.}~\bibnamefont
  {Li}}, \bibinfo {author} {\bibfnamefont {E.}~\bibnamefont {Briand}}, \bibinfo
  {author} {\bibfnamefont {C.}~\bibnamefont {Bachelet}}, \bibinfo {author}
  {\bibfnamefont {Y.}~\bibnamefont {Dumont}},\ and\ \bibinfo {author}
  {\bibfnamefont {A.}~\bibnamefont {Mougin}},\ }\bibfield  {title} {\bibinfo
  {title} {{Deviations from bulk behavior in TbFe(Co) thin films: Interfaces
  contribution in the biased composition}},\ }\href
  {https://doi.org/10.1103/PhysRevMaterials.2.104410} {\bibfield  {journal}
  {\bibinfo  {journal} {Physical Review Materials}\ }\textbf {\bibinfo {volume}
  {2}},\ \bibinfo {pages} {104410} (\bibinfo {year} {2018})}\BibitemShut
  {NoStop}%
\bibitem [{Note2()}]{Note2}%
  \BibitemOpen
  \bibinfo {note} {If $P_1$ and $P_2$ have the same sign, $v=0$ occurs at
  $L_S=-\beta L_\alpha $. If the signs are opposite, it occurs at $\beta _1 P_1
  t_1+\beta _2 P_2 t_2=0$ since $\beta _1$ and $\beta _2$ are
  positive.}\BibitemShut {Stop}%
\bibitem [{\citenamefont {Okuno}\ \emph {et~al.}(2019)\citenamefont {Okuno},
  \citenamefont {Kim}, \citenamefont {Oh}, \citenamefont {Kim}, \citenamefont
  {Hirata}, \citenamefont {Nishimura}, \citenamefont {Ham}, \citenamefont
  {Futakawa}, \citenamefont {Yoshikawa}, \citenamefont {Tsukamoto},
  \citenamefont {Tserkovnyak}, \citenamefont {Shiota}, \citenamefont
  {Moriyama}, \citenamefont {Kim}, \citenamefont {Lee},\ and\ \citenamefont
  {Ono}}]{Okuno2019}%
  \BibitemOpen
  \bibfield  {author} {\bibinfo {author} {\bibfnamefont {T.}~\bibnamefont
  {Okuno}}, \bibinfo {author} {\bibfnamefont {D.-H.}\ \bibnamefont {Kim}},
  \bibinfo {author} {\bibfnamefont {S.-H.}\ \bibnamefont {Oh}}, \bibinfo
  {author} {\bibfnamefont {S.~K.}\ \bibnamefont {Kim}}, \bibinfo {author}
  {\bibfnamefont {Y.}~\bibnamefont {Hirata}}, \bibinfo {author} {\bibfnamefont
  {T.}~\bibnamefont {Nishimura}}, \bibinfo {author} {\bibfnamefont {W.~S.}\
  \bibnamefont {Ham}}, \bibinfo {author} {\bibfnamefont {Y.}~\bibnamefont
  {Futakawa}}, \bibinfo {author} {\bibfnamefont {H.}~\bibnamefont {Yoshikawa}},
  \bibinfo {author} {\bibfnamefont {A.}~\bibnamefont {Tsukamoto}}, \bibinfo
  {author} {\bibfnamefont {Y.}~\bibnamefont {Tserkovnyak}}, \bibinfo {author}
  {\bibfnamefont {Y.}~\bibnamefont {Shiota}}, \bibinfo {author} {\bibfnamefont
  {T.}~\bibnamefont {Moriyama}}, \bibinfo {author} {\bibfnamefont {K.-J.}\
  \bibnamefont {Kim}}, \bibinfo {author} {\bibfnamefont {K.-J.}\ \bibnamefont
  {Lee}},\ and\ \bibinfo {author} {\bibfnamefont {T.}~\bibnamefont {Ono}},\
  }\bibfield  {title} {\bibinfo {title} {{Spin-transfer torques for domain wall
  motion in antiferromagnetically coupled ferrimagnets}},\ }\href
  {https://doi.org/10.1038/s41928-019-0303-5} {\bibfield  {journal} {\bibinfo
  {journal} {Nature Electronics}\ }\textbf {\bibinfo {volume} {2}},\ \bibinfo
  {pages} {389} (\bibinfo {year} {2019})}\BibitemShut {NoStop}%
\bibitem [{\citenamefont {Martínez}\ \emph {et~al.}(2019)\citenamefont
  {Martínez}, \citenamefont {Raposo},\ and\ \citenamefont
  {Alejos}}]{Martinez2019}%
  \BibitemOpen
  \bibfield  {author} {\bibinfo {author} {\bibfnamefont {E.}~\bibnamefont
  {Martínez}}, \bibinfo {author} {\bibfnamefont {V.}~\bibnamefont {Raposo}},\
  and\ \bibinfo {author} {\bibfnamefont {{\'{O}}.}~\bibnamefont {Alejos}},\
  }\bibfield  {title} {\bibinfo {title} {{Current-driven domain wall dynamics
  in ferrimagnets: Micromagnetic approach and collective coordinates model}},\
  }\href {https://doi.org/10.1016/j.jmmm.2019.165545} {\bibfield  {journal}
  {\bibinfo  {journal} {Journal of Magnetism and Magnetic Materials}\ }\textbf
  {\bibinfo {volume} {491}},\ \bibinfo {pages} {165545} (\bibinfo {year}
  {2019})}\BibitemShut {NoStop}%
\bibitem [{\citenamefont {Gambino}\ and\ \citenamefont
  {Cuomo}(1978)}]{Gambino1978}%
  \BibitemOpen
  \bibfield  {author} {\bibinfo {author} {\bibfnamefont {R.~J.}\ \bibnamefont
  {Gambino}}\ and\ \bibinfo {author} {\bibfnamefont {J.~J.}\ \bibnamefont
  {Cuomo}},\ }\bibfield  {title} {\bibinfo {title} {{Selective
  Resputtering-Induced Anisotropy In Amorphous Films}},\ }\href
  {https://doi.org/10.1116/1.569574} {\bibfield  {journal} {\bibinfo  {journal}
  {J Vac Sci Technol}\ }\textbf {\bibinfo {volume} {15}},\ \bibinfo {pages}
  {296} (\bibinfo {year} {1978})}\BibitemShut {NoStop}%
\bibitem [{\citenamefont {Hebler}\ \emph {et~al.}(2016)\citenamefont {Hebler},
  \citenamefont {Hassdenteufel}, \citenamefont {Reinhardt}, \citenamefont
  {Karl},\ and\ \citenamefont {Albrecht}}]{Hebler2016}%
  \BibitemOpen
  \bibfield  {author} {\bibinfo {author} {\bibfnamefont {B.}~\bibnamefont
  {Hebler}}, \bibinfo {author} {\bibfnamefont {A.}~\bibnamefont
  {Hassdenteufel}}, \bibinfo {author} {\bibfnamefont {P.}~\bibnamefont
  {Reinhardt}}, \bibinfo {author} {\bibfnamefont {H.}~\bibnamefont {Karl}},\
  and\ \bibinfo {author} {\bibfnamefont {M.}~\bibnamefont {Albrecht}},\
  }\bibfield  {title} {\bibinfo {title} {{Ferrimagnetic Tb–Fe Alloy Thin
  Films: Composition and Thickness Dependence of Magnetic Properties and
  All-Optical Switching}},\ }\href {https://doi.org/10.3389/fmats.2016.00008}
  {\bibfield  {journal} {\bibinfo  {journal} {Front. Mater.}\ }\textbf
  {\bibinfo {volume} {3}},\ \bibinfo {pages} {8} (\bibinfo {year}
  {2016})}\BibitemShut {NoStop}%
\bibitem [{\citenamefont {Yang}\ \emph {et~al.}(2019)\citenamefont {Yang},
  \citenamefont {Garg},\ and\ \citenamefont {Parkin}}]{Yang2019}%
  \BibitemOpen
  \bibfield  {author} {\bibinfo {author} {\bibfnamefont {S.~H.}\ \bibnamefont
  {Yang}}, \bibinfo {author} {\bibfnamefont {C.}~\bibnamefont {Garg}},\ and\
  \bibinfo {author} {\bibfnamefont {S.~S.}\ \bibnamefont {Parkin}},\ }\bibfield
   {title} {\bibinfo {title} {{Chiral exchange drag and chirality oscillations
  in synthetic antiferromagnets}},\ }\href
  {https://doi.org/10.1038/s41567-019-0438-3} {\bibfield  {journal} {\bibinfo
  {journal} {Nature Physics}\ }\textbf {\bibinfo {volume} {15}},\ \bibinfo
  {pages} {543} (\bibinfo {year} {2019})}\BibitemShut {NoStop}%
\bibitem [{\citenamefont {Chauleau}\ \emph {et~al.}(2010)\citenamefont
  {Chauleau}, \citenamefont {Weil}, \citenamefont {Thiaville},\ and\
  \citenamefont {Miltat}}]{Chauleau2010}%
  \BibitemOpen
  \bibfield  {author} {\bibinfo {author} {\bibfnamefont {J.-Y.}\ \bibnamefont
  {Chauleau}}, \bibinfo {author} {\bibfnamefont {R.}~\bibnamefont {Weil}},
  \bibinfo {author} {\bibfnamefont {A.}~\bibnamefont {Thiaville}},\ and\
  \bibinfo {author} {\bibfnamefont {J.}~\bibnamefont {Miltat}},\ }\bibfield
  {title} {\bibinfo {title} {{Magnetic domain walls displacement: Automotion
  versus spin-transfer torque}},\ }\href@noop {} {\bibfield  {journal}
  {\bibinfo  {journal} {Physical Review B}\ }\textbf {\bibinfo {volume} {82}},\
  \bibinfo {pages} {214414} (\bibinfo {year} {2010})}\BibitemShut {NoStop}%
\bibitem [{\citenamefont {Krishnia}\ \emph {et~al.}(2017)\citenamefont
  {Krishnia}, \citenamefont {Sethi}, \citenamefont {Gan}, \citenamefont
  {Kholid}, \citenamefont {Purnama}, \citenamefont {Ramu}, \citenamefont
  {Herng}, \citenamefont {Ding},\ and\ \citenamefont {Lew}}]{Krishnia2017}%
  \BibitemOpen
  \bibfield  {author} {\bibinfo {author} {\bibfnamefont {S.}~\bibnamefont
  {Krishnia}}, \bibinfo {author} {\bibfnamefont {P.}~\bibnamefont {Sethi}},
  \bibinfo {author} {\bibfnamefont {W.~L.}\ \bibnamefont {Gan}}, \bibinfo
  {author} {\bibfnamefont {F.~N.}\ \bibnamefont {Kholid}}, \bibinfo {author}
  {\bibfnamefont {I.}~\bibnamefont {Purnama}}, \bibinfo {author} {\bibfnamefont
  {M.}~\bibnamefont {Ramu}}, \bibinfo {author} {\bibfnamefont {T.~S.}\
  \bibnamefont {Herng}}, \bibinfo {author} {\bibfnamefont {J.}~\bibnamefont
  {Ding}},\ and\ \bibinfo {author} {\bibfnamefont {W.~S.}\ \bibnamefont
  {Lew}},\ }\bibfield  {title} {\bibinfo {title} {{Role of RKKY torque on
  domain wall motion in synthetic antiferromagnetic nanowires with opposite
  spin Hall angles}},\ }\href {https://doi.org/10.1038/s41598-017-11733-9}
  {\bibfield  {journal} {\bibinfo  {journal} {Scientific Reports}\ }\textbf
  {\bibinfo {volume} {7}},\ \bibinfo {pages} {11715} (\bibinfo {year}
  {2017})}\BibitemShut {NoStop}%
\bibitem [{\citenamefont {Tarasenko}\ \emph {et~al.}(1998)\citenamefont
  {Tarasenko}, \citenamefont {Stankiewicz}, \citenamefont {Tarasenko},\ and\
  \citenamefont {Ferr{\'{e}}}}]{Tarasenko1998}%
  \BibitemOpen
  \bibfield  {author} {\bibinfo {author} {\bibfnamefont {S.}~\bibnamefont
  {Tarasenko}}, \bibinfo {author} {\bibfnamefont {A.}~\bibnamefont
  {Stankiewicz}}, \bibinfo {author} {\bibfnamefont {V.}~\bibnamefont
  {Tarasenko}},\ and\ \bibinfo {author} {\bibfnamefont {J.}~\bibnamefont
  {Ferr{\'{e}}}},\ }\bibfield  {title} {\bibinfo {title} {{Bloch wall dynamics
  in ultrathin ferromagnetic films}},\ }\href
  {https://doi.org/10.1016/S0304-8853(98)00230-3} {\bibfield  {journal}
  {\bibinfo  {journal} {Journal of Magnetism and Magnetic Materials}\ }\textbf
  {\bibinfo {volume} {189}},\ \bibinfo {pages} {19} (\bibinfo {year}
  {1998})}\BibitemShut {NoStop}%
\end{thebibliography}%

\end{document}